\documentclass[traditabstract]{aa}
\usepackage{graphicx}
\usepackage{txfonts}
\usepackage{subfig}             % handles subfloats
\usepackage{epstopdf}           % .eps to .pdf

\def\ltsima{$\; \buildrel < \over \sim \;$}
\def\simlt{\lower.5ex\hbox{\ltsima}}
\def\gtsima{$\; \buildrel > \over \sim \;$}
\def\simgt{\lower.5ex\hbox{\gtsima}}
\def\cgs{{erg cm$^{-2}$ s$^{-1}$}}
\def\ergs{{erg s$^{-1}$}}
\def\cm2{{cm$^{-2}$}}

\def\xddof{{$\chi^{2}/$d.o.f.}}

\def\fhx{{$F_{\rm 2-10}$}}

\def\lum{{$L_{\rm 2-10}$}}

\def\p1{{Paper I}}

\def\xmm{{\em XMM--Newton}}

\def\asca{{\em ASCA}}
\def\nhgal{{N$_{\rm H}^{\rm Gal}$}}

\def\rosat{{\em ROSAT}}
\def\xmm{{\em XMM--Newton}}
\def\nhgal{{$N_{\rm H}^{\rm Gal}$}}
\def\nh{{$N_{\rm H}$}}

\def\f14{{10$^{-14}$}}
\def\f13{{10$^{-13}$}}
\def\f12{{10$^{-12}$}}
\def\f11{{10$^{-11}$}}

\def\4u{{4U~1344$-$60}}

\def\feka{{Fe K$\alpha$}}

\def\lbol{{$L_{\rm bol}$}}

\def\mbh{$M_{\rm BH}$}
\def\edd{$\lambda_{\rm Edd}$}
\def\ledd{$L_{\rm Edd}$}
\def\nus{{\em NuSTAR}}
\def\ecut{$E_{\rm cut}$}
\def\swift{{\it Swift}}

\usepackage{color}
\usepackage{amstext}

\begin{document} 

\title{\nus\ reveals the extreme properties of the super-Eddington accreting SMBH in PG 1247+267}
   \author{G. Lanzuisi\inst{1,2}
   \and
   M. Perna\inst{1,2}
   \and
   A. Comastri\inst{2}
   \and
   M. Cappi\inst{3} 
   \and   
   M. Dadina\inst{3}
   \and
   A. Marinucci\inst{4}
   \and 
   A. Masini\inst{1,2}
   \and
   G. Matt\inst{4}
   \and
   F. Vagnetti\inst{5}
   \and
   C. Vignali\inst{1,2}
   \and
   D. R. Ballantyne\inst{6}
   \and
   F. E. Bauer\inst{7,8,9}
   \and
   S. E. Boggs\inst{10}
   \and
   W. N. Brandt\inst{11,12,13}
   \and
   M. Brusa\inst{1,2}
   \and
   F. E. Christensen\inst{14}  
   \and
   W. W. Craig\inst{10,15}     
   \and
   A. C. Fabian\inst{16}     
   \and
   D. Farrah\inst{17}   
   \and
   C. J. Hailey\inst{18}
   \and
   F. A. Harrison\inst{19}
   \and  
   B. Luo\inst{20,21}
   \and
   E. Piconcelli\inst{22}
   \and
   S. Puccetti\inst{23,22}
   \and
   C. Ricci\inst{7}
   \and
   C. Saez\inst{24}
   \and
   D. Stern\inst{25}
   \and    
   D. J. Walton\inst{19,25}
   \and  
   W.W. Zhang\inst{26}
}

\titlerunning{Super-Eddington accretion in PG 1247+267 as observed by \nus}\authorrunning{G.~Lanzuisi et al.}

\institute{
Dipartimento di Fisica e Astronomia, Universit\`a  di Bologna, Viale Berti Pichat 6/2, I--40127 Bologna, Italy \and                                       
INAF - Osservatorio Astronomico di Bologna,  Via Ranzani 1, I--40127 Bologna, Italy \and                                                                 
INAF - Istituto di Astrofisica Spaziale e Fisica Cosmica, Via Piero Gobetti 101, I--40129 Bologna, Italy \and                                            
Dipartimento di Matematica e Fisica, Universit\`a di Roma Tre, Via della Vasca Navale 84, I--00146 Roma, Italy \and                                      
Dipartimento di Fisica, Universit\`a di Roma Tor Vergata, Via della Ricerca Scientifica 1, I--00133 Roma, Italy \and                                      
Center for Relativistic Astrophysics, School of Physics, Georgia Institute of Technology, 837 State St., Atlanta, GA 30332, USA\and   
Instituto de Astrof\'{\i}sica, Facultad de F\'{i}sica, Pontificia Universidad Cat\'{o}lica de Chile, Casilla 306, Santiago 22, Chile\and  
Millennium Institute of Astrophysics, Vicuna Mackenna 4860, 7820436 Macul, Santiago, Chile \and                                                          
Space Science Institute, 4750 Walnut Street, Suite 205, Boulder, CO 80301, USA \and 
Space Science Laboratory, University of California, Berkeley, CA 94720, USA \and                                                                         
Department of Astronomy and Astrophysics, 525 Davey Lab, The Pennsylvania State University, University Park, PA 16802, USA \and                          
Institute for Gravitation and the Cosmos, The Pennsylvania State University, University Park, PA 16802, USA \and                                         
Department of Physics, 104 Davey Lab, The Pennsylvania State University, University Park, PA 16802, USA \and                                                                                                                  
DTU Space National Space Institute, Technical University of Denmark, Elektrovej 327, 2800 Lyngby, Denmark \and                                           
Lawrence Livermore National Laboratory, Livermore, CA 94550, USA \and                                                                 
Institute of Astronomy, Madingley Road, Cambridge CB3 0HA, UK \and
Department of Physics, Virginia Tech, Blacksburg, VA 24061, USA  \and
Columbia Astrophysics Laboratory, Columbia University, New York, NY 10027, USA \and                                                                      
Cahill Center for Astronomy and Astrophysics, California Institute of Technology, Pasadena, CA 91125, USA \and                                           
School of Astronomy and Space Science, Nanjing University, Nanjing, 210093, China \and
Key laboratory of Modern Astronomy and Astrophysics (Nanjing University), Ministry of Education, Nanjing 210093, China \and
INAF - Osservatorio Astronomico di Roma, Via di Frascati, 33, I-00040 Monteporzio Catone (Roma), Italy\and             
ASDC--ASI, Via del Politecnico, 00133 Roma, Italy \and
Korea Astronomy and Space Science Institute, 776 Daedeokdae-ro, Yuseong-gu, Daejeon 305348, Korea \and
Jet Propulsion Laboratory, California Institute of Technology, Pasadena, CA 91109, USA  \and                                                             
NASA Goddard Space Flight Center, Greenbelt, MD 20771, USA                                                                                               
}

\date{Received 17 February 2016; Accepted 21 March 2016}

\abstract
{

PG1247+267 is one of the most luminous known quasars at $z\sim2$ and is a strongly super-Eddington accreting SMBH candidate.
We obtained \nus\ data of this intriguing source in December 2014 with the aim of studying its high-energy emission, 
leveraging the broad band covered by the new \nus\ and the archival \xmm\ data.
Several measurements are in agreement with the super-Eddington scenario for PG1247+267: the soft power law ($\Gamma=2.3\pm0.1$);
the weak ionized Fe emission line; and a hint of the presence of outflowing ionized gas surrounding the SMBH.
The presence of an extreme reflection component is instead at odds with the high accretion rate proposed for this quasar. 
This can be explained with three different scenarios; all of them are in good agreement with the 
existing data, but imply very different conclusions:
i) a variable primary power law observed in a low state, superimposed on a reflection component echoing a past, higher flux state; 
ii) a power law continuum obscured by an ionized, Compton thick, partial covering absorber; and 
iii) a relativistic disk reflector in a lamp-post geometry, with low coronal height and high BH spin.
The first model is able to explain the high reflection component in terms of variability. 
The second does not require any reflection to reproduce the hard emission, while a rather low high-energy cutoff of $\sim100$~keV is detected
for the first time in such a high redshift source.  
The third model require a face-on geometry, which may affect the SMBH mass and Eddington ratio measurements.
Deeper X-ray broad-band data are required in order to distinguish between these possibilities.

}{}{}{}{}


\maketitle

\section{Introduction}
The Eddington luminosity (expressed as $L_{\rm Edd}=4\pi G M m_p c/\sigma_T $) is an approximate upper bound to the total luminosity \lbol\
that can be radiated by a compact object of mass $M$, set by the equilibrium between the radiation pressure acting outward
and the gravitational force acting inward, for a spherically symmetric geometry (Eddington 1916). 
The limit of \lbol/\ledd\ (hereafter \edd) = 1 is thought to regulate the growth of supermassive black holes (SMBHs) 
over cosmic time and is observed to hold both at low  (e.g., Schulze \& Wisotzki 2010) and high redshift (e.g., Nobuta et al. 2012, Suh et al. 2015).
However, super-Eddington accretion periods are thought to be possible (Zubovas \& King 2013) and may even be required 
to explain the fast growth of the first SMBHs (see, e.g., Volonteri 2012 for a review); 
super-Eddington episodes could in fact be the transient phenomena in which SMBHs 
gain most of their mass (King 2003).

As implied by its rapid variability, the X-ray emission is produced in the inner regions surrounding the SMBH 
(few to tens of gravitational radii, $r_{\rm G}$; e.g., Chartas et al. 2009, Zoghbi et al. 2012; Reis \& Miller 2013)
and is the most direct probe for investigating accretion properties.
There are several key observables in the X-ray spectra of active galactic nuclei (AGN) related to the accretion rate.
First, a positive correlation between spectral slope ($\Gamma$) and flux (and hence \edd)
is thought to be a common feature of SMBH accretion and has been observed  in individual, highly variable AGN at low redshift
(e.g., Perola et al. 1986, Vaughan \& Edelson 2001); in large samples of AGN, both locally and at high redshift 
(Shemmer et al. 2008; Risaliti et al. 2009, Brightman et al. 2013); 
and also derived theoretically for BHs in general (Laurent \& Titarchuk 2011),
with sources close to the Eddington limit showing softer spectra ($\Gamma$ up to 2.5) compared to sources accreting 
at slower rates ($\Gamma\sim1.5$). There is also evidence that this relation tends to flatten above \edd$=1$ (Shih et al. 2002, Ai et al. 2011).
The relation between $\Gamma$ and \edd\ is based on the existence of a strong link between the accretion flow and the properties of both
the disk and the corona: a high accretion rate drives the disk temperature up, producing more soft X-ray radiation and at
the same time increases the Compton cooling of the corona, steepening the slope of the X-ray continuum (e.g., Shemmer et al. 2006).

Second, an anti-correlation has been observed between \edd\ and the equivalent width (EW) of the narrow component of the Fe K$\alpha$
emission line (EW$\propto$ \edd$^{-0.19}$; Bianchi et al. 2007): 
at \edd=1 the line can be as weak as EW = 25 eV.
This correlation is thought to be the fundamental cause underlying the observed {\it X-ray Baldwin effect} 
(Iwasawa \& Taniguchi 1993), in which the EW anti-correlates with the X-ray luminosity. 
The nature of this correlation and the role of selection effects in this context, however, are still debated (see Shu et al. 2012).

The $\Gamma$-\edd\ and EW-\edd\ relations may be related; objects with higher values of \edd\ have a steeper continuum, 
implying that the number of photons that can produce the fluorescent Fe K$\alpha$ emission is smaller, leading to a smaller 
EW (Ricci et al. 2013). 
The intensity of the fluorescent emission line and the Compton hump are also related
since for a given reflector geometry, ionization parameter, and Fe abundance the relative strength of the Fe
K$\alpha$ line and the reflected continuum is fixed (e.g., George \& Fabian 1991); therefore, 
in highly accreting sources the last component should also be weaker.
Furthermore, AGN with high \edd, both at low and high redshift, tend to have stronger ionized \feka\ emission at 6.9 keV, 
with respect to the neutral \feka\ line at 6.4 keV (Nandra et al. 1996; Iwasawa et al. 2012).
The same is observed in narrow-line Seyfert 1 galaxies (e.g., Comastri et al. 1998) that are also thought to have high \edd.
This is probably due to the fact that the inner disc is more strongly ionized when the central source is brighter (see, e.g., Keek \&  Ballantyne 2015).

Finally, gas outflows are naturally expected during super-Eddington accretion phases (Zubovas \& King 2012)
owing to the intense radiation pressure associated with these events.
Such winds are one of the main candidates thought to produce the AGN ``feedback'' and to fine-tune the \mbh$-\sigma$ relation observed in the 
local Universe between quiescent SMBHs and their host bulges (see Kormendy \& Ho 2013 for a review).
These outflows are observed as blueshifted absorption features produced by ionized gas in optical/UV spectra of $\sim20$\% of optically selected quasars
(e.g., Weymann et al. 1979, Gibson et al. 2009), and are also observed  in 30-40\% of X-ray spectra of local AGN (Tombesi et al. 2010) 
and in a few high redshift QSOs (Chartas et al. 2003, Saez et al. 2011, Lanzuisi et al. 2012, Vignali et al. 2015) in the form of highly ionized Fe absorption lines.
These absorbers may be part of a single large-scale, 
stratified outflow observed at different locations from the black hole and 
spanning several orders of magnitude in ionization, column density, and velocity (Tombesi et al. 2013). 
Considering the few QSOs at low redshift that show strong signatures of massive winds and have reliable BH mass measurements, 
all appear to be accreting close to their Eddington limit:
PG1211+143, PG0844+349, PDS 456, MCG-6-30-15 (Pounds et al. 2003a, 2003b, Reeves et al. 2003).

The luminous QSO PG1247+267 (hereafter PG1247) at $z=2.048$ is candidate to be an extreme example of super-Eddington accretion,
and in this paper we present new \nus\ observations obtained with the aim of characterizing its high-energy emission.
The source properties are summarized in Sect.~2. The existing \xmm\ data (Sec.~3) are combined with the 
hard X-ray data from \nus\ (Sec.~4) to constrain the high-energy properties of this extreme QSO (Sec.~5).
Results from the broad-band X-ray spectral fits are summarized and discussed in Sec.~6.
Throughout the paper, a standard $\Lambda-$CDM cosmology with $H_0=70$ km s$^{-1}$ Mpc$^{-1}$, $\Omega_\Lambda=0.7$, and $\Omega_M=0.3$ is used.
Errors are quoted at the 90\% confidence level for one parameter of interest.

\section{Source properties}

%======================================================
 \begin{figure}[t]
 \begin{center}
 \includegraphics[width=7.5cm]{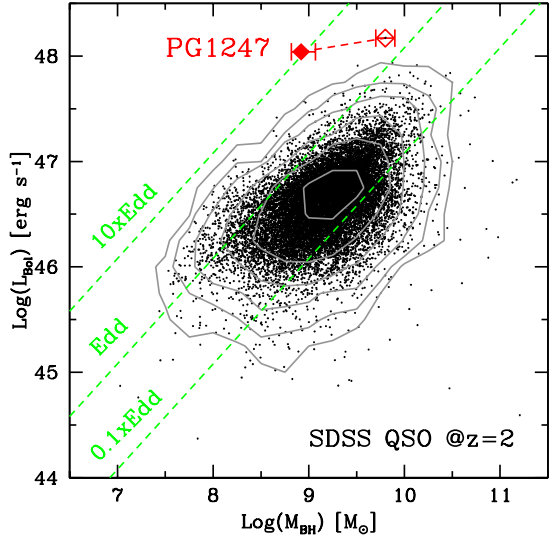}\vspace{0.5cm}
 \includegraphics[width=5.5cm]{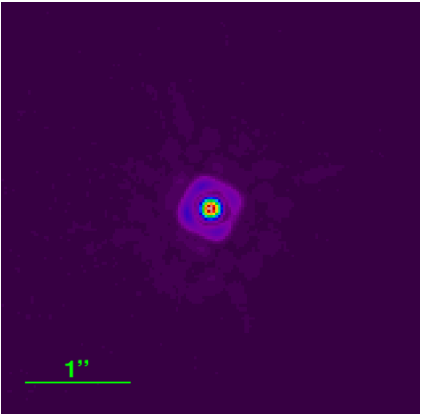}
 \caption{{\it Top:} \lbol\ and \mbh\ for the $1.8<z<2.2$ SDSS QSO sample (black points and gray contours).
 Values for PG1247 from SDSS  (Shen et al. 2011) and from Trevese et al. (2014) are shown with empty and filled red diamonds, respectively.
 Dashed green lines mark different \edd\ values.
 {\it Bottom:} $4\arcsec\times4\arcsec$ {\it HST} NICMOS image of PG1247 (filter F160W). The 1\arcsec\ scale is shown.
 }
 \label{fig:selection}
 \end{center}
 \end{figure}
%=======================================================

Having \lbol$=1.5\times10^{48}$ \ergs, an absolute magnitude of $M_i~\sim~-30$, and an observed magnitude of $i=15.37$, 
PG1247 is the most luminous radio-quiet\footnote{The source 
is undetected (at $5\sigma$ c.l.) in the FIRST radio survey at 1.4 GHz, and has a radio loudness $R<0.3$ (defined as $R=f_{1.4{\rm GHz}}/f_{4400\AA}$).}
QSO at $z\sim2\pm0.2$ in the SDSS DR7 (Abazajian et al. 2009, Shen et al. 2011)\footnote{The second considering SDSS 1521+5202 at $z=2.208$ with $M_i~\sim~-30.6$, Luo et al. (2015).}.
PG1247 is also the second most luminous AGN with a reverberation mapping (RM) BH mass measurement available, 
obtained from a spectro-photometric monitoring campaign with the 1.82 m Copernicus Telescope at Asiago (Trevese et al. 2007),
after the lensed quasar APM 08279+5255 at $z\sim4$ (Saturni et al. 2015). 
The BH mass estimate from C{\scriptsize III]} and C{\scriptsize IV} RM is \mbh$=8.3^{+3.4}_{-2.7}\times10^8$ $M_{\odot}$ (Trevese et al. 2014). 
The average \lbol\ obtained from the optical continuum (between 2500\AA\ and  5100\AA) measured in these observations is $1.1\times10^{48}$ \ergs.
These values translate into a fiducial value of \edd$=10.6$, the highest reported in the literature to date for a QSO with a SMBH mass obtained
via RM.

The source also has  a single epoch (SE) BH mass estimate, reported in Shen et al. (2011), from Mg{\scriptsize II} in the range Log(\mbh/$M_{\odot}$)$=9.7-9.9$,
depending on the calibration adopted,
and Log(\mbh/$M_{\odot}$)$\sim9.8$ from C{\scriptsize IV}. 
The authors take the latter as the \mbh\ fiducial value for all the sources with $z>1.9$ because of the SDSS observing band limits.  
Although the statistical errors on these measurements are small ($\Delta$\mbh$=0.01-0.03$)
one must bear in mind that there are large systematic uncertainties (\simgt$0.4$ dex) associated with these estimates.
In particular, SE estimates based on C{\scriptsize IV} are known to be affected by large uncertainties due to the presence of outflowing components 
contributing to the C{\scriptsize IV} line flux (see, e.g., the discussion in Shen et al. 2011).
The strong blue wing in the C{\scriptsize IV} emission line and the presence of absorption systems observed in PG1247 SDSS spectrum support this possibility. 
In particular, a doublet absorption line at $\sim$ 1508$\AA$ with a rest-frame separation of 2.57$\AA$, FWHM $\sim200$ km/s and EW doublet ratio of $\sim0.8$ 
identify the 1548 and 1550 transitions of the C{\scriptsize IV} doublet line (see, e.g., Vestergaard 2003). 
The doublet is at $\sim-7700$ km s$^{-1}$ from the systemic and therefore can be associated with outflowing gas.

On the other hand, the RM measurements based on the same emission line
are less affected by  non-virial outflows  if the root mean square spectrum (Peterson et al. 1998) is used instead of the average spectrum, 
because the outflow components are expected to vary on timescales
longer than reverberation time lags (see the discussion in Trevese et al. 2014).
In all cases, even taking into account the SE value, the source has an accretion rate in the range \edd$=0.9-1.75$, which is close to or above the Eddington limit.
All these properties are summarized in Fig.~1 (top), where the \lbol\ and \mbh\ values for PG1247 from Trevese et al. (2014) and 
from Shen et al. (2011) are shown in comparison with the $1.8<z<2.2$ quasars from the SDSS DR7 catalog.
PG1247 is a truly remarkable source that can be used as a laboratory for testing predictions of super-Eddington accretion theories.

A magnification factor of $\mu=5-10$ due to strong lensing could bring the Eddington ratio of PG1247 to $\sim1$ (see discussion in Trevese et al. 2014).
However, the quasar was observed several times between 1991 and 1998 with {\it HST} 
as a single point source, with no sign of multiple images, down to scales of $\sim0.1$\arcsec.
Figure~1 (bottom) shows a $4\arcsec\times4\arcsec$ zoom around PG1247 from the combined {\it HST}-NICMOS observations of 1998 (F160W filter, total exposure time 1150s).
The possibility of strong lensing is therefore excluded, as all known high-redshift lensed QSOs observed so far with {\it HST}
have been resolved into multiple images at those scales.\footnote{See, e.g., the CASTLES database of {\it HST} gravitational lenses (https://www.cfa.harvard.edu/castles/)
observed with comparable exposure times.}

\section{Existing X-ray data}
%======================================================
 \begin{figure}[t]
 \begin{center}
 \vspace{0.5cm}
 \includegraphics[width=7.5cm]{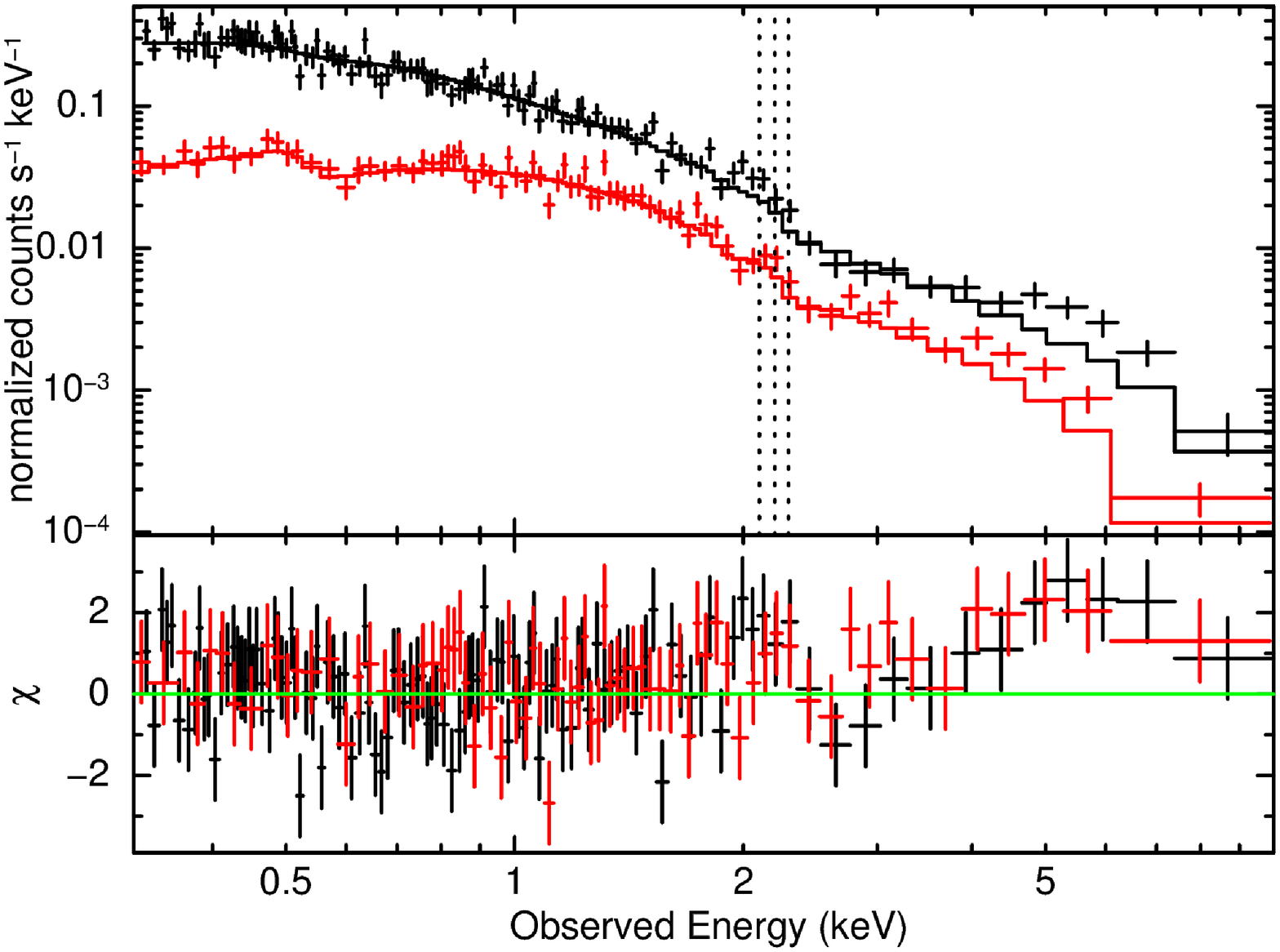}\vspace{0.5cm}
 \includegraphics[width=7.5cm]{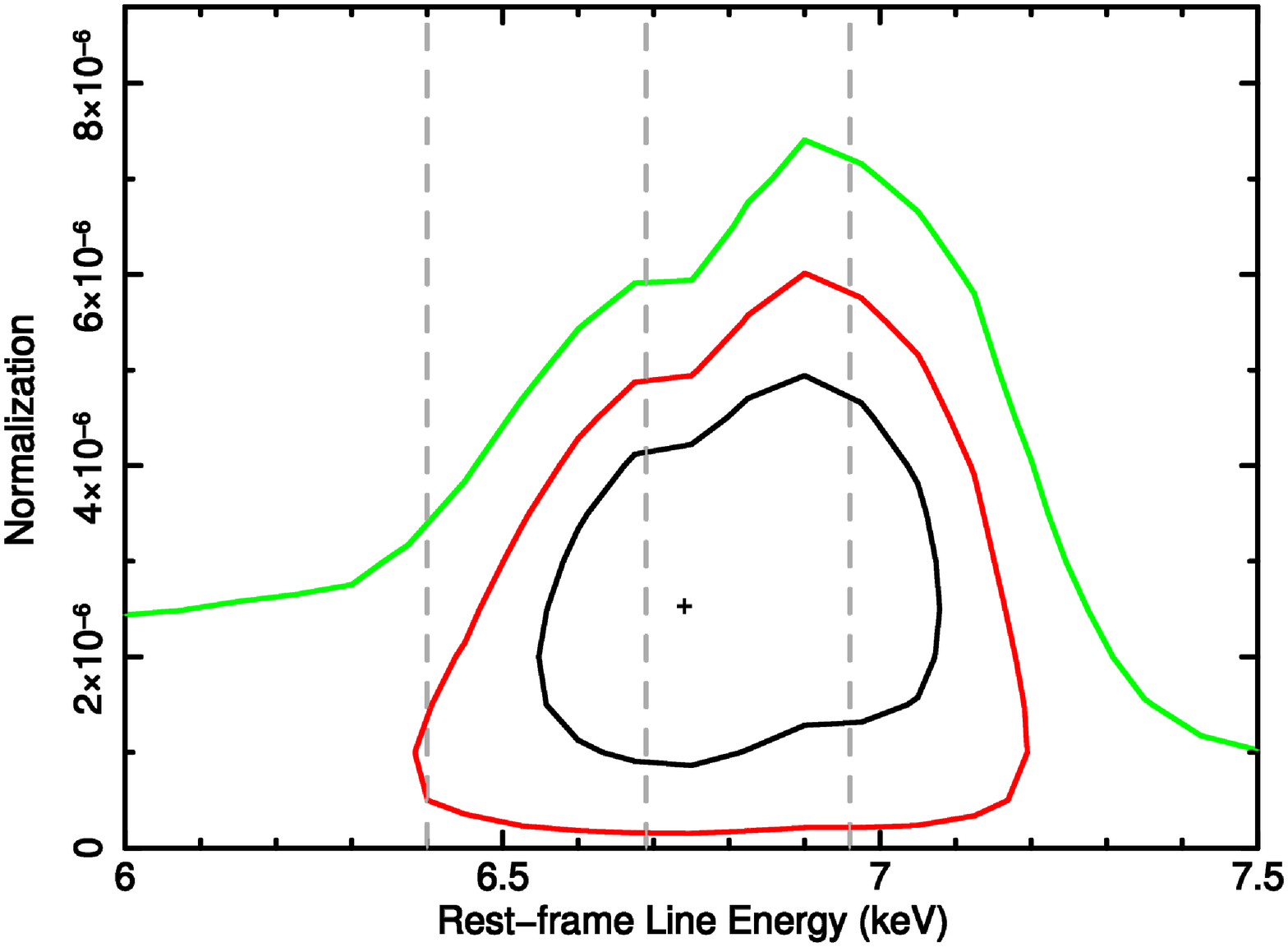}
 \caption{{\it Top}: \xmm\ pn (black) and MOS1+2 (red) spectra of PG1247. The model is a simple power law 
 fitted to the 0.3-2 keV observed band to highlight the residuals above 2 keV due to the emission line complex
 and the strong reflection component.
 {\it Bottom}: Contour plot (at 68, 90, and 99\% c.l.) for the emission line energy (rest frame) and normalization.
 In both panels the dashed lines mark the expected neutral Fe, FeXXV, and FeXXVI K$\alpha$ emission line energies.}   
 \label{fig:specxmm}
 \end{center}
 \end{figure}
%=======================================================

PG1247 was observed by \xmm\ in 2003 for a total of 34~ks (20 ks after background-flare subtraction).
No significant variability is detected within the \xmm\ observation.
Page et al. (2004) analyzed the data, and reported a strong
reflection component ($R\sim2.8$, modeled with {\it PEXRAV}, Magdziarz \& Zdziarski 1995), a very soft power law ($\Gamma=2.23\pm0.10$),
plus a broad and neutral \feka\ line ($\sigma=0.52\pm0.35$ keV, $EW=421\pm215$~eV).
As the authors pointed out, such a high reflection fraction is unphysical in the context of the {\it PEXRAV} model, 
which includes a primary power law and models reflection from cold material in a slab geometry,
since $R$ is defined to be $\Omega/2\pi$, i.e., the solid angle of the cold material visible from the 
Comptonizing source in units of $2\pi$, and therefore $R\sim2.8>4\pi$ sr, the solid angle of the entire sphere. 

We reanalyzed the data applying standard extraction procedures using the Scientific Analysis System (SAS) v13.5,
and optimized extraction regions (40\arcsec\ for pn and 35\arcsec\ for MOS1 and 2) to obtain pn and MOS1+2 spectra\footnote{We verified that the source is too faint to have useful RGS data.}. 
The extracted spectra have $\sim4560$ pn and $\sim2900$ MOS1+2 counts in the 0.3-10 keV band (black and red data points in Fig.~2, top). 
We fitted the spectrum with a single power law plus Galactic absorption (\nhgal$=0.9\times10^{20}$ cm$^{-2}$, Kalberla et al. 2005)
in the observed $0.3-2$ keV band, corresponding to the rest frame $1-6$ keV band
at the source redshift. Indeed, extrapolating this model to the full 0.3-10 keV observed band, strong residuals can be seen 
around the expected position of the \feka\ emission line, and at the reflection hump. 
We fitted the data with the same {\it PEXRAV} reflection model adopted in Page et al. (2004) to check for consistency.
The best fit (\xddof$=278.5/300$) is obtained with a very soft power law with $\Gamma=2.36\pm0.08$,
again with an extremely high reflection component $R=3.92_{-1.25}^{+1.80}$.
The high-energy cutoff (\ecut) is fixed to 100 keV in the \xmm\ data, given that the \xmm\ spectra only reach $\sim30$ keV 
(rest frame), and are therefore not able to constrain this parameter, given the available photon statistics.
Also, the inclination angle  (defined as the inclination angle
between the line of sight and the vertical axis of the accretion disk) is fixed to the default value, $\cos(i)=0.45$ 
(see Section 5.1.1. for the effect of possible different geometries).
The normalizations of pn and MOS1+2 are left free to vary independently, but they are in agreement within $\sim4$\%.

Residuals to the {\it PEXRAV} model in both the pn and MOS1+2 spectra around 6.4 keV rest frame 
suggest the presence of an emission line close to the rest-frame energies of the \feka\ line.
We added  a narrow ($\sigma=10$ eV fixed) emission line to the model at energy $E_{\rm line}=6.4$~keV, fixed.
The improvement in the fit is small ($\Delta \chi^2$=1.9)
and the addition of the emission line is justified only at a $\sim85\%$ confidence level (c.l., F-test probability $p=0.15$).\footnote{
The F-test is known to yield only an approximate probability of finding an emission line at a given energy, and only
if the line energy is known in advance (see, e.g., Protassov et al. 2002, Markowitz et al. 2006).}
Leaving the energy of the line as a free parameter, we obtain a further improvement of the fit of $\Delta \chi^2$=3.0.
The addition of the new free parameter is justified at $\sim93\%$ c.l. ($p=0.071$ from an F-test).
The best-fit line energy is $6.78_{-0.32}^{+0.42}$ keV.
Fig.~2 (bottom) shows the confidence contours of the emission-line energy and normalization, obtained
for a fixed $\sigma=10$~eV. 
As can be seen in the contours, the observed feature could be the result of blending
of Fe XXV and Fe XXVI, while the 6.4 keV energy is ruled out at 90\% c.l.
Leaving the line width $\sigma$ as a free parameter, the best-fit values become $E_{\rm line}=6.82_{-0.30}^{+0.42}$~keV
and $\sigma\leq0.63$~keV. The improvement in the fit is minimal (\xddof$=272.6/297$) and the new free parameter is not justified ($p=0.30$).
As pointed out in Page et al. (2004), although there is evidence that the iron line may be broadened, 
no definite statement can be made.
A narrow emission line with the same parameters derived above is included in all the following fits.
The equivalent width of the line is EW$=168_{-150}^{+198}$~eV rest frame.
Given the inferred luminosity of \lum$=10^{46}$ \ergs, the X-ray Baldwin effect predicts $EW=20-30$~eV (Bianchi et al. 2007).
Assuming that this value refers to a canonical, $R=1$ reflection component with solar abundances,
the observed emission line EW is broadly consistent with the strong reflection component ($R=3-4$) found above.

PG1247 was also observed by \swift\ XRT 19 times between 2007 and 2014. We collected the spectra in all the observations
with more than 1 ks exposure time (14 in total), and merged them  using standard {\it ftool} procedures to obtain a single spectrum
with statistics good enough to perform a basic spectral analysis. The final spectrum has $\sim560$ net counts in the 0.3-10 keV
band for a total exposure time of 43 ks. 
When fitted with the same reflection model described above, the XRT data give consistent results with the \xmm\ spectra:
a steep power law ($\Gamma=2.2\pm0.4$) and a strong reflection component ($R\sim2.5$).
Given the limited number of counts, and the fact that they have been collected over 7 years, we decided not to use the XRT
spectrum for the following spectral analysis. However, we used this  best-fit model to convert the observed count 
rate of each observation into full band (0.2-10 keV) fluxes, which will be relevant for the variability analysis of the source (see section 5.2 and Fig.~7).

\section{\nus\ data}

 %======================================================
 \begin{figure}[t]
 \begin{center}
 \includegraphics[width=8.5cm]{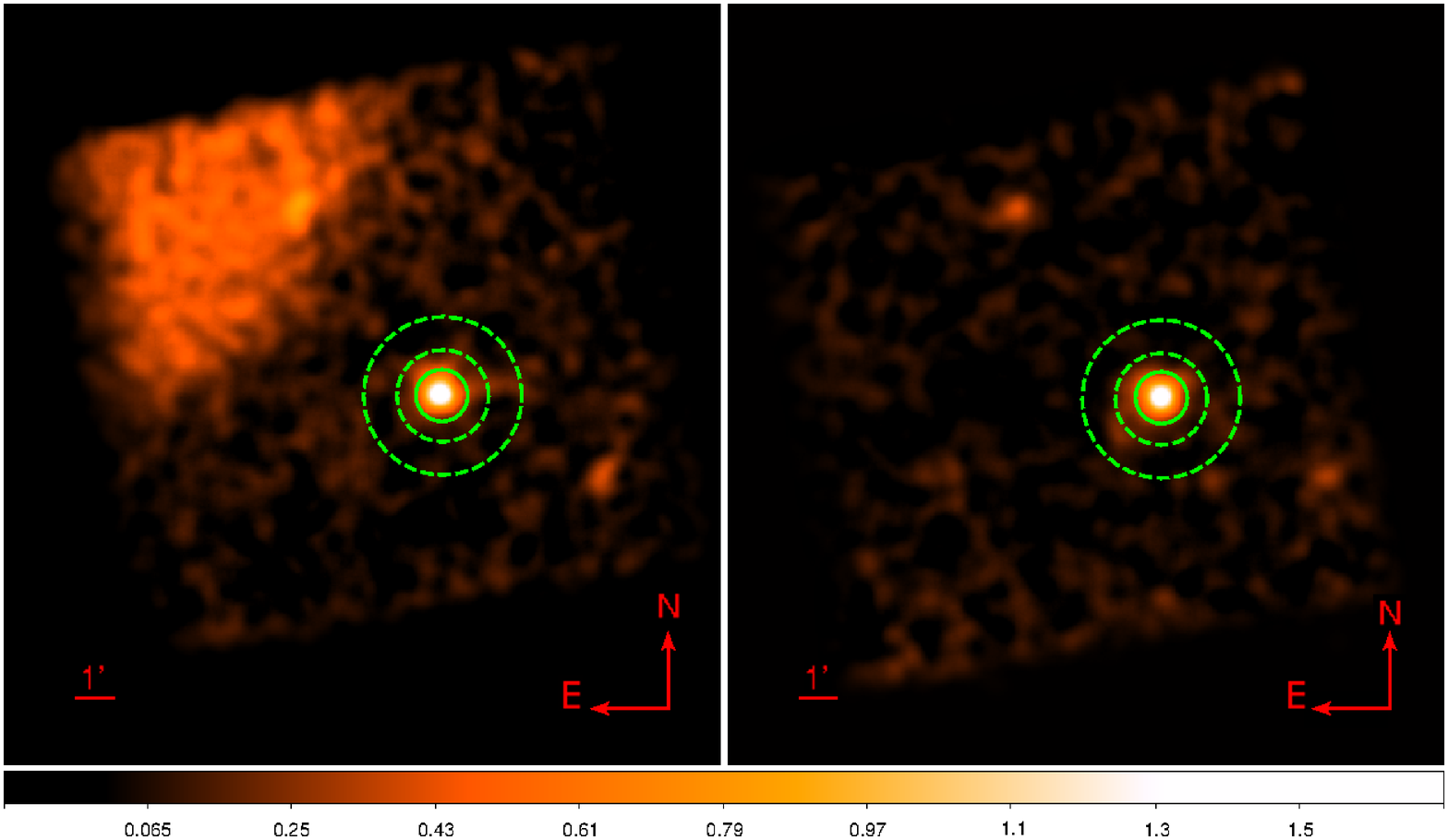}\hspace{0.5cm}
  \includegraphics[width=8.5cm]{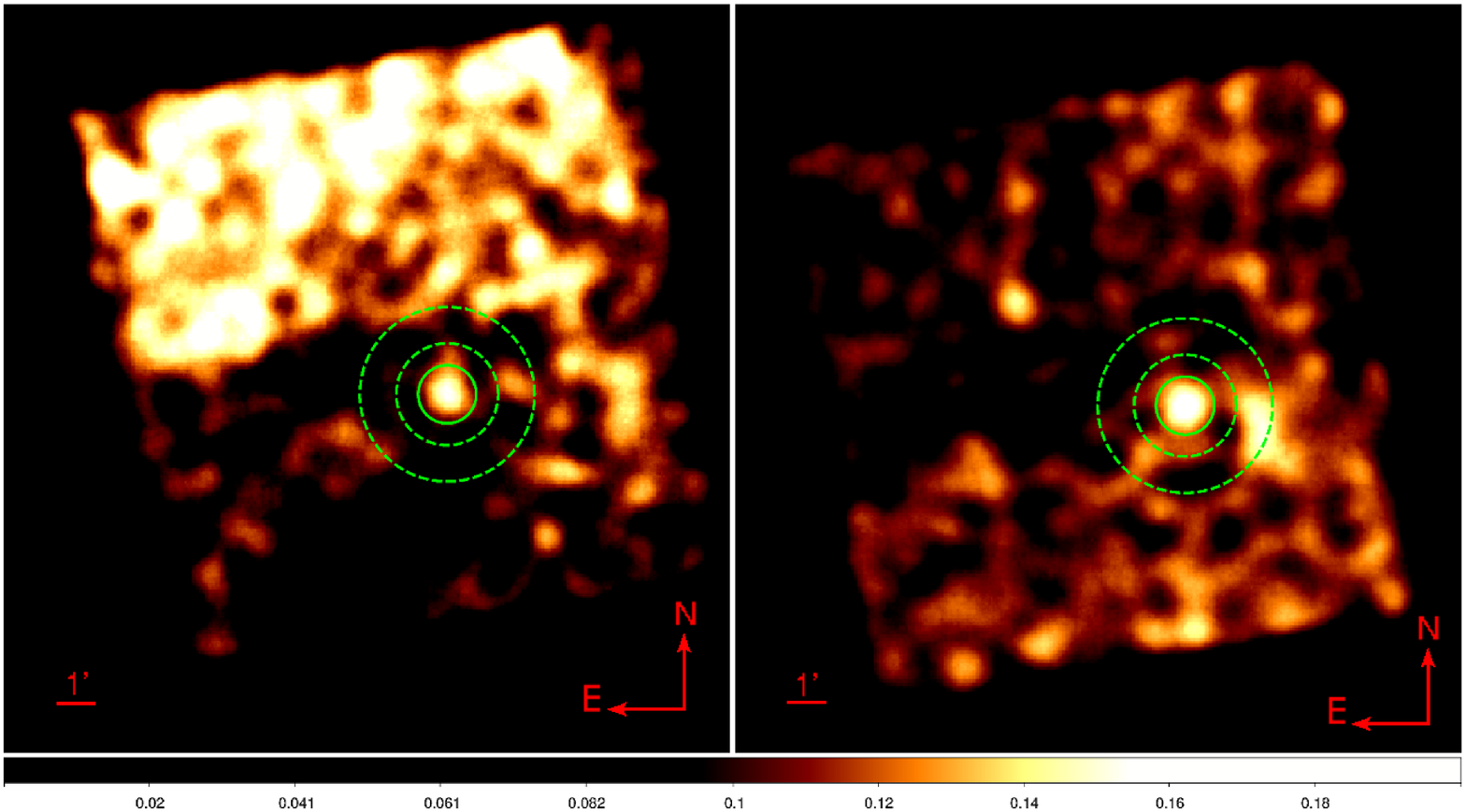}
 \caption{{\it Top:} 3-79 keV background-subtracted images of the \nus\ observation of PG1247 (FPMA left and FPMB right). 
  Contamination from the Coma cluster is visible in the upper left corner of FPMA.
 {\it Bottom:} 15-30 keV \nus\ images, not background subtracted.  
 In both panels the $1\arcmin$ scale is shown, and the source and background extraction regions are shown with continuous and dashed lines, respectively. North is up and east is left. 
 }
 \label{fig:nuskybkgimages}
 \end{center}
 \end{figure}
%=======================================================
 
%======================================================
 \begin{figure}[t]
 \begin{center}
 \vspace{0.5cm} 
 \includegraphics[width=7.5cm]{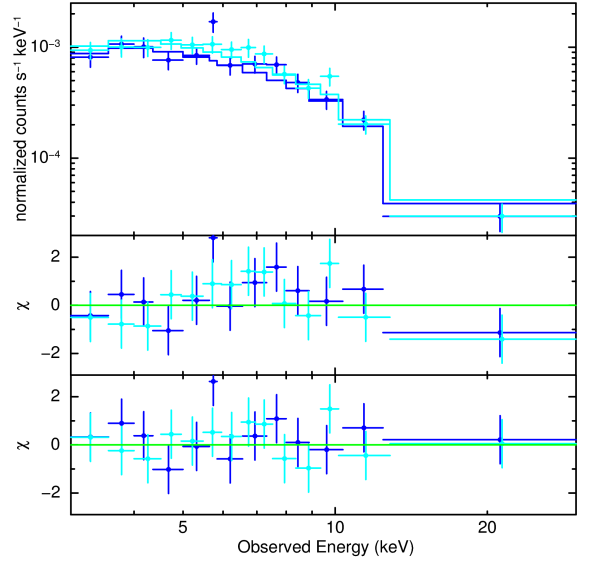}
 \caption{\nus\ FPMA and FPMB data (blue and cyan points, respectively) fitted with a simple power law (top panel). 
 The spectra have been further rebinned with respect to the $S/N=3$ level to highlight the curvature of the spectrum in the residuals (central panel).
 The residuals with respect to the fit performed with the {\it PEXRAV} model are shown in the lower panel.}
 \label{fig:specnustar}
 \end{center}
 \end{figure}
%=======================================================

PG1247 was observed by \nus\ (Harrison et al. 2013) in December 2014 for a total of 94 ks.
The raw data were processed using the \nus\ Data
Analysis Software package v. 1.4.1 (NuSTARDAS)\footnote{http://heasarc.gsfc.nasa.gov/docs/nustar/analysis/nustar\_swguide.pdf}. 
Calibrated and cleaned event files were produced using the calibration files
in the \nus\ CALDB (version 20150312) and standard filtering criteria
with the {\it nupipeline} task. 
The global light-curves extracted from the full FPMA and FPMB modules in the energy range 3-20 keV
show a strong background flare --- with a count-rate more than a factor of two higher than the stable level ---
due to solar activity, lasting $\sim2$ hours. 
We removed this high background period using the {\it nustar$\_$filter$\_$lightcurve} IDL script. The cleaned exposure time is 86 ks.

The \nus\ observation of PG1247 is also affected by contamination (particularly in FPMA)
by the Coma cluster, lying $\sim2.5^{\circ}$ northeast of the quasar.
Fig. 3 (top), shows the 3-79 keV images of PG1247 obtained with the FPMA (left) and FPMB (right) detectors
produced with the IDL code  {\it nuskybkg} (Wik et al. 2014). This code takes into account 
all the background components usually observed in \nus, i.e., the instrumental background, the focused cosmic X-ray background, and the aperture background.
However, the code does not model and remove the contamination from the nearby cluster, which is evident in the upper left corner of the FPMA image. 
Therefore, we used annular regions to extract local background spectra instead of using the background files produced by {\it nuskybkg}.
The background spectra were extracted from annular regions of inner radius 70\arcsec\ and outer radius 120\arcsec.
Circular extraction regions for the source, with different radii, were tested in order to find the 
radius that optimizes the $S/N$ for each focal plane module. The final extraction radius is 40\arcsec\
for both FPMA and FPMB, which gives a $S/N$ of $\sim$14 and $\sim$15, respectively.
The {\it nuproducts} task was then used to extract the \nus\ source and background spectra, plus the appropriate response and ancillary files.
In the following the fits are performed leaving the relative normalizations between FPMA and FPMB as free parameters.
The two normalizations agree within 15\% and are consistent within $1\sigma$.

%=======================================================
\begin{figure*}[t]
\begin{center}
\includegraphics[width=7cm]{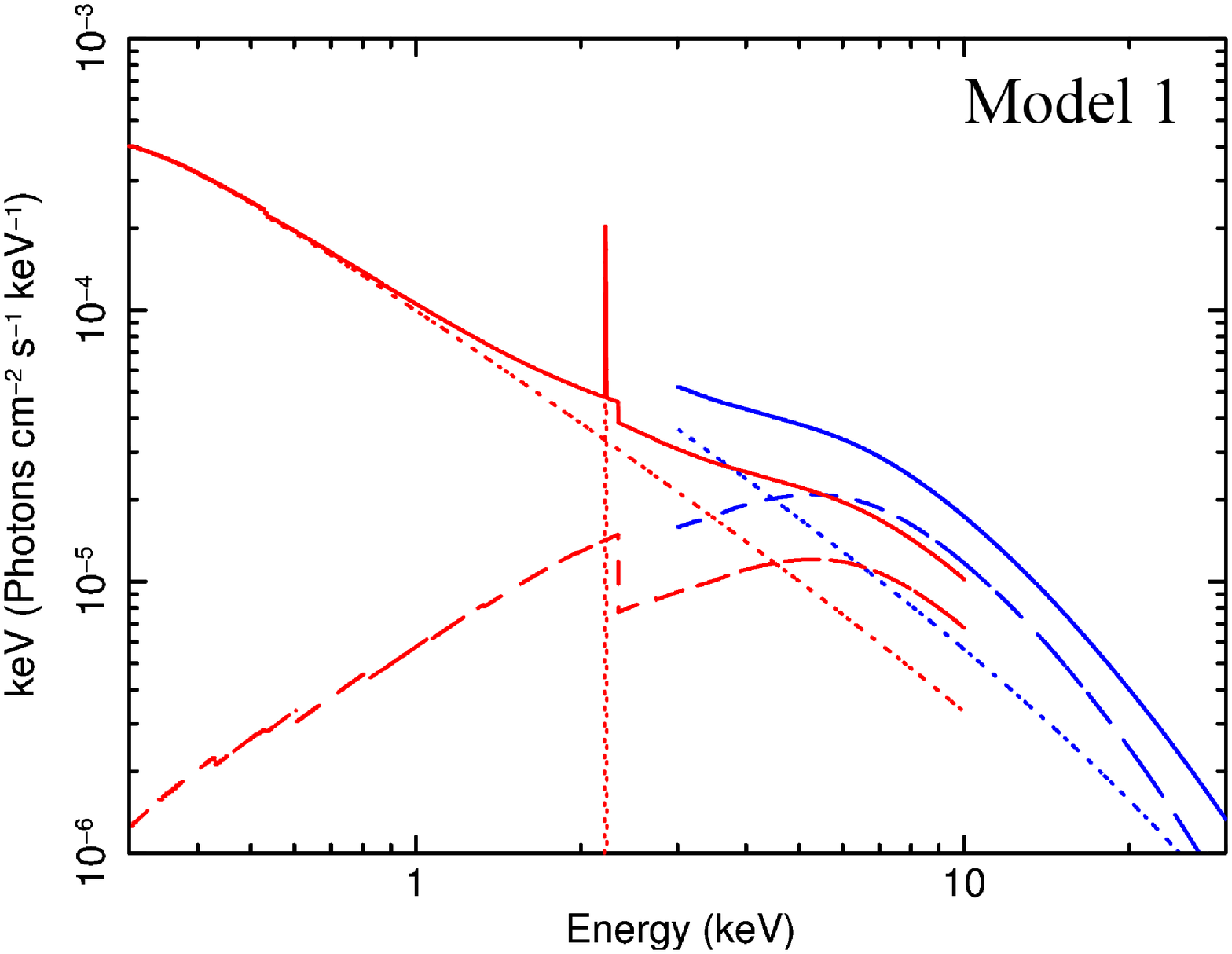}\hspace{0.5cm}\includegraphics[width=7cm]{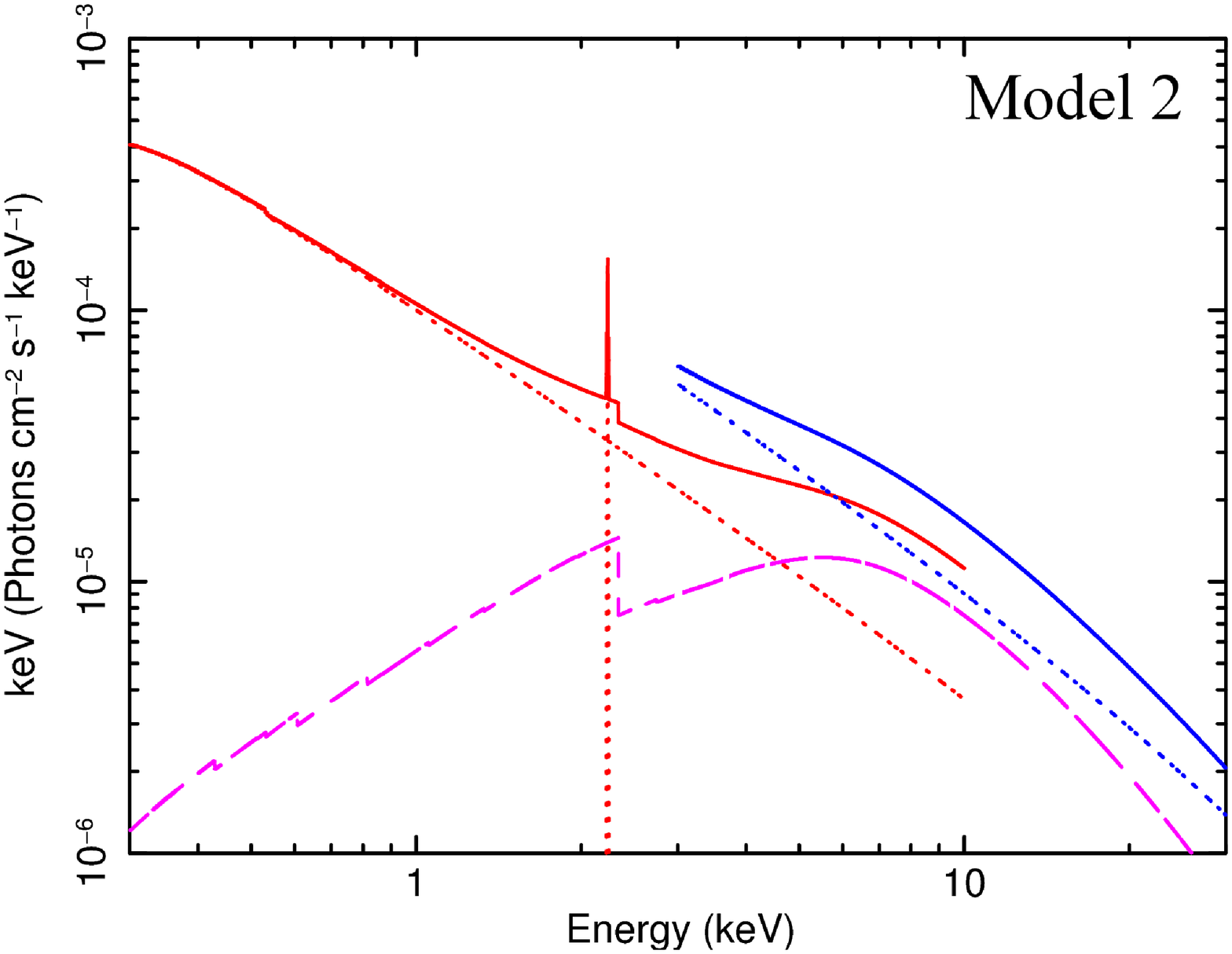} \vspace{0.5cm}
\includegraphics[width=7cm]{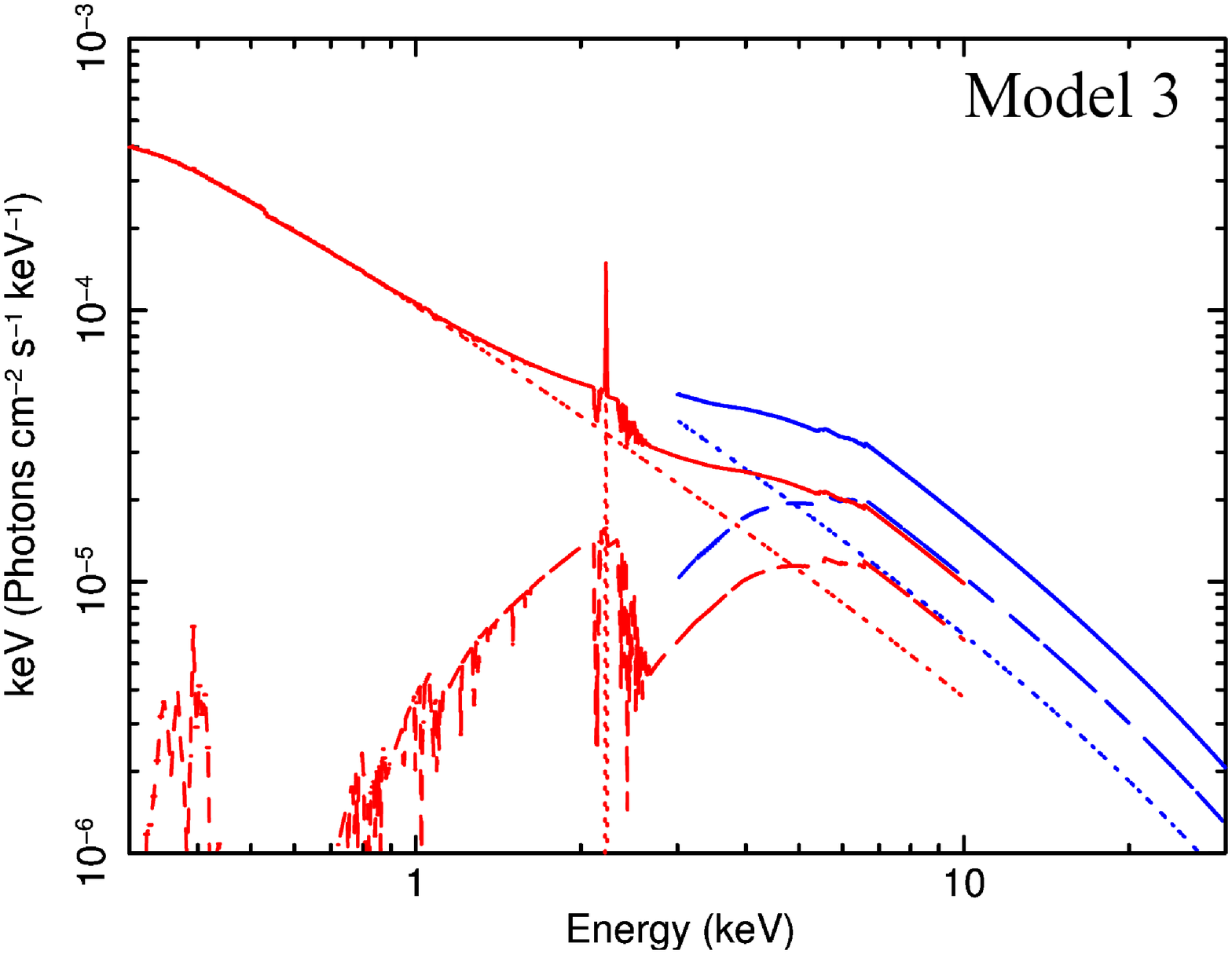}\hspace{0.5cm}\includegraphics[width=7cm]{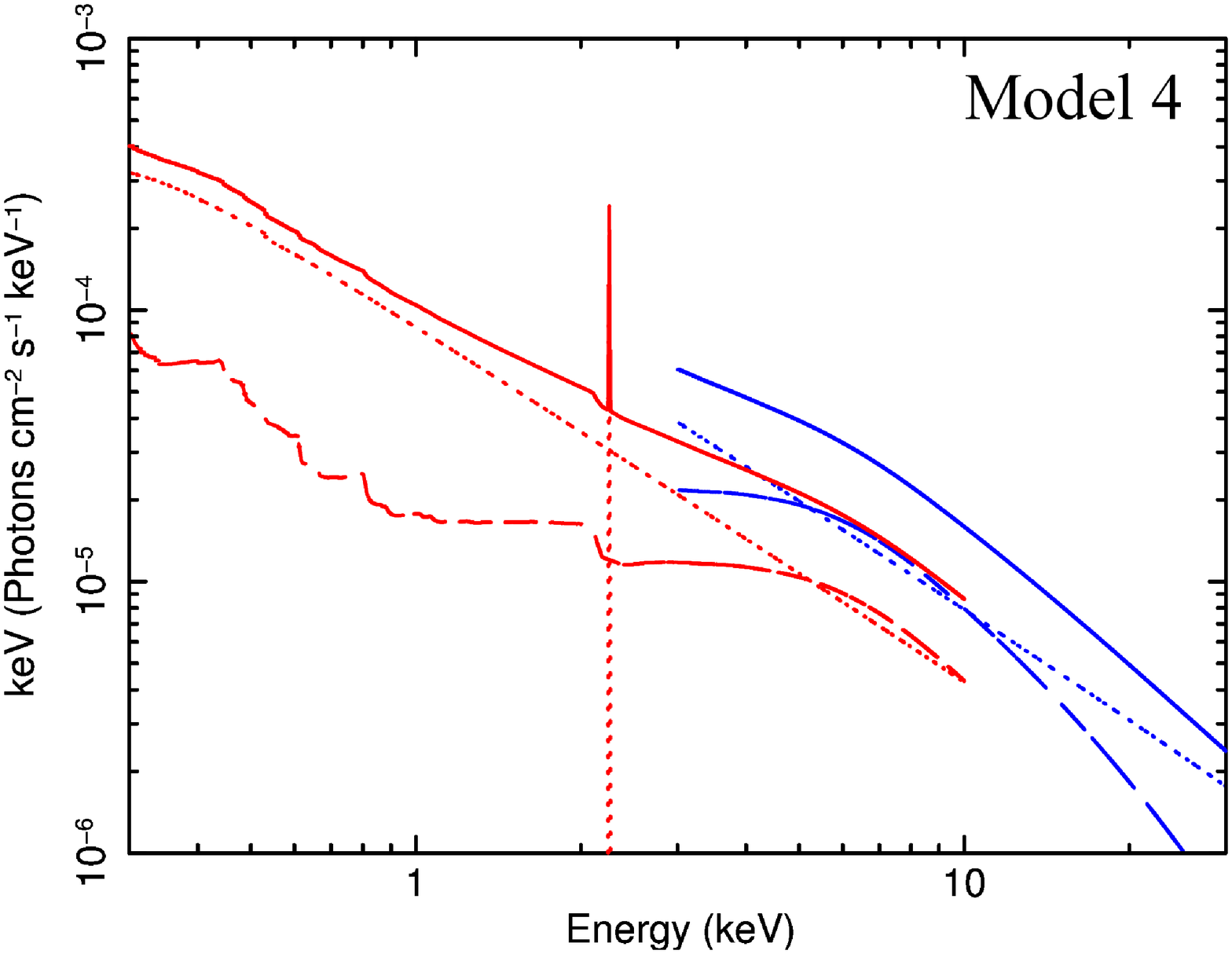}
\caption{Best-fit unfolded models for models 1, 2, 3, and 4, from top left to bottom right. The continuous curves represents the best fit
to the \xmm\ data (in red) and \nus\ data (in blue). The dotted curves represent the primary power law,
while the dashed curves represent the component reproducing the hump (either reflection or absorption).
In the top right panel the magenta curve represents the constant reflection component underlying both \xmm\ and \nus\ data, modeled with {\it PEXRAV}.
}
\label{fig:models}
\end{center}
\end{figure*}
%=======================================================

The spectra have been grouped using the tool {\it Specgroup} within SAS, to a minimum $S/N=3$ per bin.
The source is detected at this level between 3 and 30 keV in both instruments,
with a total of 520 (590) net counts for FPMA (FPMB).
Fig.~3 (bottom) shows the 15-30 keV \nus\ image of PG1247 without any background subtraction: the source is clearly detected even in this very hard band.  
Fig.~4 shows the \nus\ spectra (FPMA in cyan and FPMB in blue) fitted with a simple power law ($\Gamma=2.14\pm0.13$, \xddof$=65.46/73$). 
The curved shape of the hard X-ray spectrum of PG1247 is visible in the residuals (central panel).
Adopting a {\it PEXRAV} model, the fit is significantly improved (\xddof$=55.83/71$). Residuals for this fit are shown in the lower panel of Fig.~4.
However,  owing to the limited photon statistics and to the complexity of the underlying model, the \nus\ spectra alone do not provide interesting simultaneous
constraints on the high-energy cutoff, 
the reflection fraction, and the photon index;
the best-fit photon index is $\Gamma=1.9\pm1.5$, while the high-energy cutoff and reflection fraction are only loosely constrained
(\ecut$>50$ keV, $R>1.35$).

The flux in the 2-10 keV band (extrapolated using the {\it PEXRAV model}) is \fhx$=(4.2\pm0.5)\times10^{-13}$ \cgs,
while the 2-10 keV flux measured by \xmm\ with the same model was \fhx$=(2.3\pm0.1)\times10^{-13}$ \cgs:
the \nus\ data show that the source has increased in flux between the \xmm\ observation in 2003 and the \nus\ observation in 2014 by a factor of $\sim1.8$.
Also, the intermediate {\it Swift}-XRT observations show an average flux level consistent with the \nus\ measurement.
The cross-calibration between \nus\ and the other X-ray satellites is accurate within 10\% (Madsen et al. 2015).

Because the correlation between $\Gamma$ and flux (i.e., \edd) saturates at high fluxes (e.g., Shih et al. 2002, Ai et al. 2011),
we try to fit the \nus\ spectra with $\Gamma$ fixed at the \xmm\ value to get more stringent constraints on the \ecut\ and $R$ parameters, 
and test whether there is any difference, for example, in the $R$ value derived from \xmm\ due to the change in flux (i.e., accretion rate).
The resulting best fit \ecut\ is \ecut$>60$~keV, while the reflection fraction is $R=2.7_{-1.4}^{+3.4}$,
and therefore consistent with the \xmm\ results within the large errors.

\section{Joint spectral modeling}

We modeled the \xmm\ and \nus\ spectra together with the aim of deriving a physically meaningful interpretation 
of the peculiar X-ray spectrum of PG1247, leveraging the broad band covered by the combined data and accounting for the higher flux measured during the \nus\ observation.
First, we adopted a cold reflection model and tested whether the addition of the new \nus\ data allow for a less extreme value of $R$ (Sec. 5.1).
Then we tested whether the strong reflection seen in \xmm\ can be explained in terms of variability of the continuum superimposed on a 
constant cold reflection component (Sec.~5.2).
Third, we tested the possibility that the hump observed in the hard X-rays is produced by absorption of partial covering, dense material, rather than by reflection (Sec.~5.3). 
Finally we tested a model in which the reflection is produced in the inner regions of an ionized accretion disk, and is relativistically blurred (Sec.~5.4).

%=======================================================
\begin{table}
\caption{Fit parameters for the baseline models discussed in Section 5.}
\begin{center}
\begin{tabular}{lccccl}
\hline\hline\\
Model 1:                     &  &  &  &  &  \texttt{pexrav} (var. cold refl.) \\
 & & & & & \\               
 $\Gamma$                    &  &  &  &  & $2.35_{-0.08}^{+0.09}$   \\
 \ecut\ ~~~~(keV)                &  &  &  &  &   $89_{-34}^{+112}$  \\
 $R$                         &  &  &  &  &  $3.8_{-1.5}^{+2.1}$     \\
 \xddof                      &  &  &  &  &   $327.1/367$             \\
 \hline\\
Model 2:                     &  &  &  &  &  \texttt{pexrav} (const. refl.) \\
 & & & & & \\                
 $\Gamma$                    &  &  &  &  &  $2.37_{-0.08}^{+0.09}$    \\
 \ecut\ ~~~~(keV)                &  &  &  &  &   $163_{-83}^{+470}$  \\
 $R$                         &  &  &  &  & $3.8\pm1.4$ / $1.7\pm0.8$ $^a$  \\
 \xddof                      &  &  &  &  &   $327.7/367$              \\
 \hline\\          
Model 3:                     &  &  &  &  &  \texttt{pexrav $\times$ zxipcf} (ion. abs.)\\
 & & & & & \\
 $\Gamma$                    &  &  &  &  &  $2.33\pm0.10$       \\
 \ecut\ ~~~~(keV)                &  &  &  &  &   $96_{-47}^{+181}$  \\
 \nh\ ~~~~~(cm$^{-2}$)            &  &  &  &  &   $1.3\pm0.3\times10^{24}$  \\
 CF                          &  &  &  &  &   $0.62\pm0.1$              \\
 $\log\xi$ ~~~(erg cm s$^{-1}$) &  &  &  &  &   $2.4\pm0.4^b$ \\
 \xddof                      &  &  &  &  &  $325.5/365$               \\
 \hline\\           
Model 4:                     &  &  &  &  &  \texttt{relxill\_lp} (relativistic refl.) \\
 & & & & & \\
 $\Gamma$                    &  &  &  &  &  $2.26\pm0.04$   \\
 $h$  ~~~~~~~~~($r_{\rm G}$)            &  &  &  &  &   $<3.5$              \\
 $a$                         &  &  &  &  &   $>0.68$                   \\
 $\cos(i)$                   &  &  &  &  &    $>0.83$                \\
 $\log\xi$ ~~~(erg cm s$^{-1}$) &  &  &  &  &   $1.8\pm0.5^c$   \\
 \xddof                      &  &  &  &  &  $333.8/365$              \\

 \hline\hline\\
\end{tabular}
\end{center} 
Model 1 is a {\it PEXRAV} model with all the parameters linked between \xmm\ and \nus, except for the normalizations.
Model 2 has the normalization of the reflection component linked between the \xmm\ and \nus\ data,
while the primary power law is left free to vary independently, so that the two data sets have two different $R$ parameters.
The $\Gamma$ and high-energy cutoff are linked between the power-law and the reflection component, and between the two data sets.
Model 3 includes an ionized absorber with partial covering superimposed on a {\it PEXRAV} model with $R=0$.
Model 4 is a {\it RELXILL\_LP} relativistic reflection model, where the \ecut\ is frozen to 300 keV. 
In Models 1, 2, and 3 the inclination angle is frozen to the default value $\cos(i)=0.45$.
$^a$ $R$ values for \xmm\ and \nus, respectively.
$^b$ Ionization parameter of the obscuring material.
$^c$ Ionization parameter of the accretion disk.
\end{table}
%=======================================================

\subsection{Variable cold reflection}
%======================================================
 \begin{figure}
 \begin{center}
 \includegraphics[width=8.5cm]{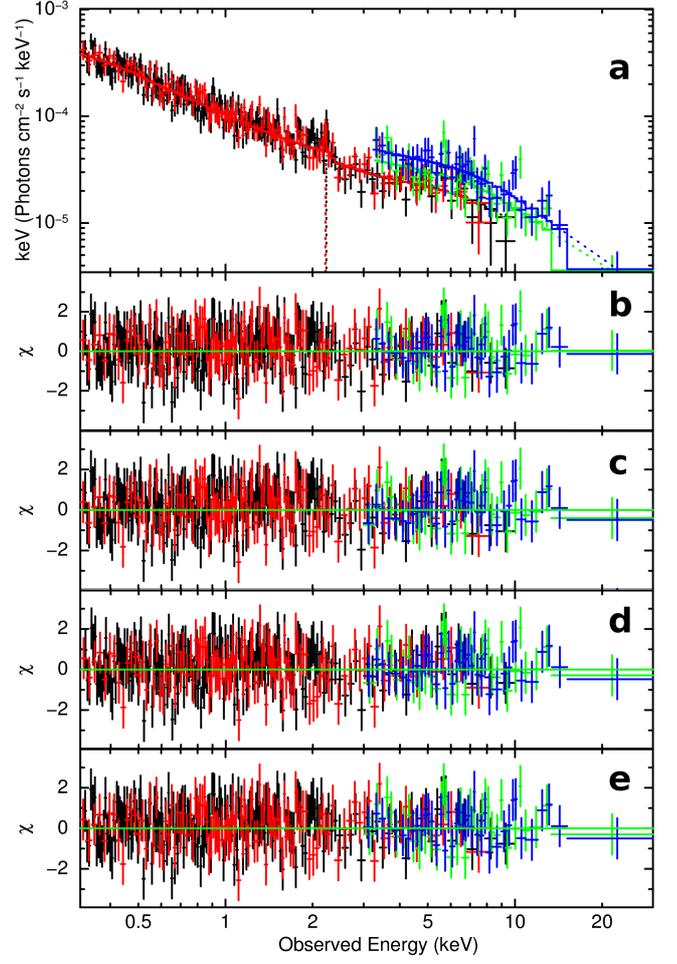}
 \caption{{\it Panel a}: Unfolded data and model for the best fit obtained with model 1. {\it Panels b, c, d, and e}: 
 delta $\chi$ for model 1, 2, 3, and 4, respectively.  
 In all panels \xmm\ pn data are in black, MOS1+2 in red, \nus\ module A in green, and B in blue.}
 \label{fig:specuf}
 \end{center}
 \end{figure}
%=======================================================
To account for the overall variability of the X-ray spectra between 2003 and 2014,  
we modeled the \xmm+\nus\ data with a cold reflection model
({\it PEXRAV}) as described in Section 3, but where the normalizations of the \xmm\ and \nus\ data are allowed to vary independently,
while the rest of the free parameters, i.e., $\Gamma$, \ecut,\ and $R$, are forced to be the same for the two data sets (model 1);
 i.e., we are assuming that the spectral shape was the same in the two periods, and only the global normalization has changed.
The inclination angle was first fixed to $\cos(i)=0.45$.

As a first step we fixed the value of the reflection parameter to $R=1$ to check whether the addition of \nus\ data allow 
for a less extreme reflection component with respect to the \xmm\ data alone.
The resulting fit gives \xddof$=347.8/369$, with a soft photon index ($\Gamma=2.22\pm0.05$), 
while \ecut\ is not well constrained.
We then left the reflection parameter $R$ free to vary (model 1, Fig.~5, top left). 
The best-fit value for the reflection parameter is $R=3.8_{-1.5}^{+2.0}$ and the improvement in the fit is large 
($\Delta\chi^2=20.1$). The F-test indicates that the addition of this extra free parameter provides a 
significant improvement to the fit (F-test probability $p=2.7\times10^{-6}$).
The best-fit photon index is again very soft ($\Gamma=2.35_{-0.08}^{+0.09}$), comparable to the one obtained from \xmm\ alone.
With this model the \ecut\ is constrained to a rather low value (\ecut$=89_{-34}^{+112}$ keV rest frame).
The best-fit parameters are given in Table 1. The unfolded data and model for the best fit obtained in this way are shown in panel (a) of Fig.~6. Panel (b) shows the $\Delta\chi$ for the same model.

We note that a small inclination angle $i$ between the line of sight and the accretion disk axis
can increase the reflection component observed by at most a factor of $\sim2$ with respect to the default geometry used in this section ($\cos(i)=0.45$).
Therefore, even an extremely face-on geometry cannot entirely account for the $R>3$ reflection observed in the data.
 We obtain $R=2.5_{-1.0}^{+1.4}$ for $\cos(i)=0.95$, still significantly larger than $R=1$, which is the reflection for a $2\pi$ sr solid angle.
We therefore discard this model as unrealistic.

%======================================================
 \begin{figure}[t]
 \begin{center}
 \includegraphics[width=7.5cm, height=6.5cm]{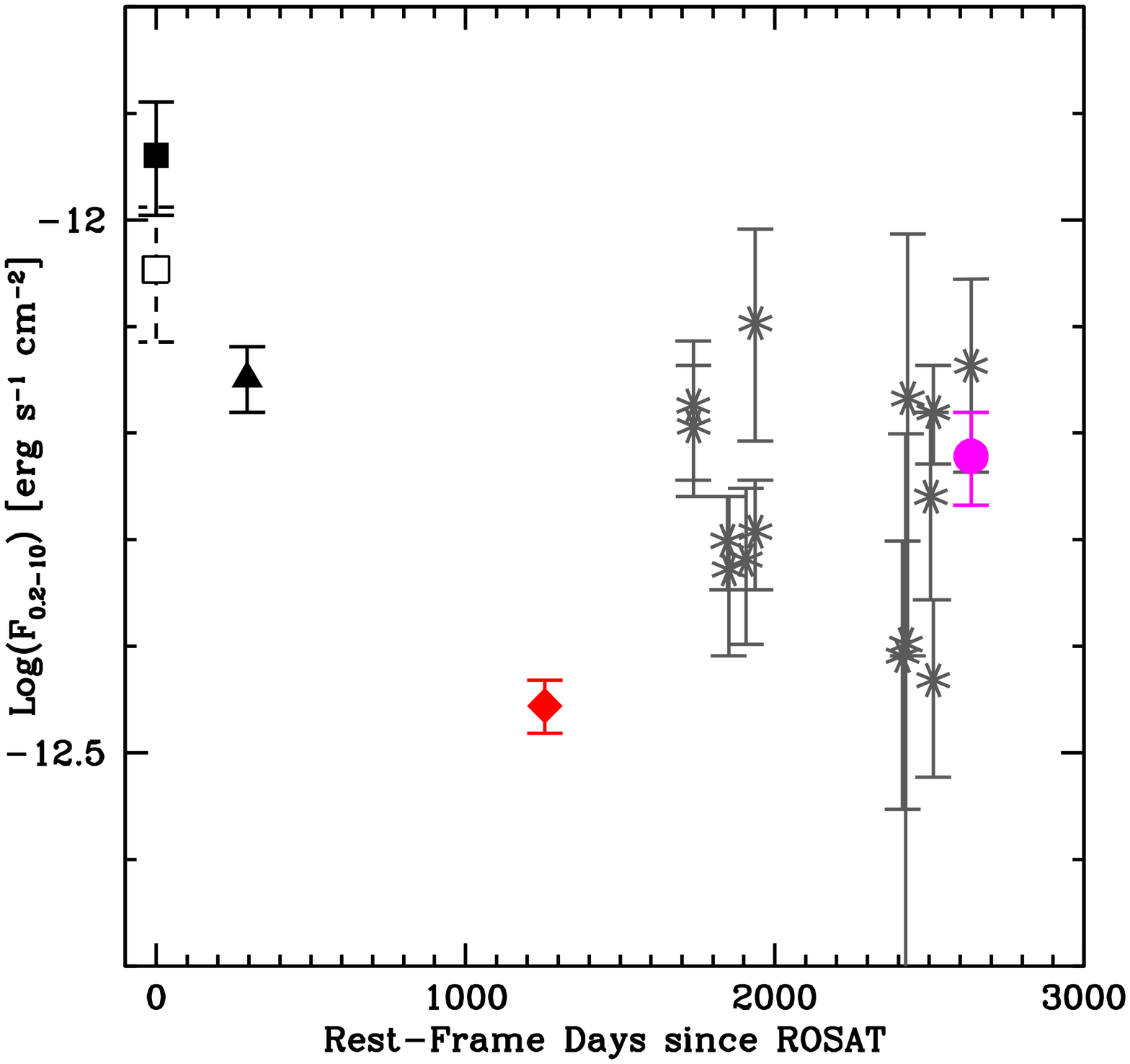}
 \caption{Rest-frame 0.2-10 keV light-curve of PG1247  since its first detection with \rosat\ in 1993 (black filled square).
 The empty square shows the \rosat\ data reanalyzed in Boller et al. (2016).
 The black triangle represents the \asca\ measurement (1995).
 The red diamond represents the \xmm\ measurement of 2003. The gray points represent the monitoring performed with \swift\ between
 2007 and 2013 (Shemmer et al. 2014),  and the magenta point shows the \nus\ measurement in 2014.}
 \label{fig:lc}
 \end{center}
 \end{figure}
%=======================================================

\subsection{Constant cold reflection}

To reproduce a case in which a variable primary power law is superimposed on a constant reflection component,
we used a model that includes a power law with high-energy exponential cutoff and the normalization left 
free to vary between the \xmm\ and \nus\ observations, plus a {\it PEXRAV} component
seen only in reflection ($R=-1$) and kept constant between the two observations (model 2, Fig.~5 top right).
The photon index and \ecut\ are left as free parameters, but are forced to have the same value in the power-law and {\it PEXRAV} components. 
In this way, the extreme $R$ parameter observed in the \xmm\ spectra of 2003
could be explained in terms of a primary power law observed in a lower state
with respect to the reflection hump, which is expected to be less variable and to exhibit significant time lags,
being the result of reflection from distant material (i.e., the torus) with a range of different light paths.
The goodness of the fit in this case is comparable with that of model 1 (\xddof=$327.7/367$).
The $\Delta\chi$ for this model are shown in panel (c) of Fig.~6.
The resulting $R$ parameters for the fit with a constant reflection component are $R=3.8\pm1.4$ for
the \xmm\ spectra and $R=1.7\pm0.8$ for the \nus\ spectra.
If this model is correct, the best-fit cutoff is \ecut $\sim$ 160~keV (rest frame), a factor of 2 higher than the value measured for model 1 and with larger errors:
it is loosely constrained to be \ecut\simlt600~keV  at 90\% c.l. Given that the \nus\ data only reach a 90~keV rest frame, some doubt can be cast on the reliability of this result.

The weak Fe line EW observed in the \xmm\ data (EW$=168_{-150}^{+198}$~eV, see Section 3) is consistent, within the errors, with the expected Fe line EW computed 
taking into account the high reflection factor observed in that data set and the X-ray Baldwin effect 
($EW\sim20-30$ eV for the luminosity of PG1247, to be rescaled by a reflection parameter $R=3.8$)
without requiring any particularly low Fe abundance.

Fig.~\ref{fig:lc} shows the 0.2-10 keV (rest-frame) light-curve of PG1247 as has been observed since 1993 by \rosat, \asca, \xmm,\ and \swift. 
The light-curve is taken from Shemmer et al. (2014). We added four more \swift\ observations taken after November 2013, 
and as an empty square the flux obtained from the reanalysis of the \rosat\ data 
performed in Boller et al. (2016), with a better background treatment. 
The \nus\ data point (magenta) is added by extrapolating the hard \hbox{X-ray} flux
to the 0.2-10 keV flux, using the spectral model described in this section.
PG1247 is highly variable at \hbox{X-ray} wavelengths despite its exceptional X-ray luminosity:
the normalized excess variance (e.g., Nandra et al. 1997) is $\sigma^2_{rms}=0.14\pm0.09$ (Shemmer et al. 2014), 
higher than expected for its X-ray luminosity, i.e., $\sigma^2_{rms}\sim0.01$ (Young et al. 2012, Lanzuisi et al. 2014).
More importantly, the light-curve shows that the flux level of the \xmm\ observation (red point) is the lowest measured in the last twenty years: 
the X-ray flux dropped from $\sim10$ to 3.5$\times10^{-13}$ erg s$^{-1}$ cm$^{-2}$
between the \rosat\ and the \xmm\ measurements from 1993 to 2003 
($\sim1200$ days rest frame), while the flux measured by \swift\ between 2007 and 2013 
spans the range 4-8$\times10^{-13}$ erg s$^{-1}$ cm$^{-2}$.
In this scenario, we obtain a lower limit of the distance between the primary X-ray source and the reflector
of $\sim1200$ light-days, equivalent to $\sim1$ pc, which is consistent with molecular torus size estimates obtained from 
the source luminosity (see Section~6).

\subsection{Ionized partial covering absorber} 

We then tested the possibility that an ionized, partial covering absorber could reproduce the shape observed in PG1247
without the need of any reflection component (or with a standard $R=1$ cold reflection) by affecting mostly the X-ray data below 10 keV (rest frame).
Model 3 (Fig.~5, bottom left) includes an ionized absorber with partial covering ({\it ZXIPCF} in Xspec\footnote{This model is based on the {\it Xstar} photo-ionization code
(Kallman \& Bautista 2001), where the ionization parameter is defined as $\xi=L/(n*R^2)$, $L$ is the luminosity of the ionizing source, 
$n$ the density of the ionized medium, and $R$ the distance between the two.}, Reeves et al. 2008), 
superimposed on a {\it PEXRAV} model with the reflection fraction parameter set to $R=0$ (i.e., no reflection).
Indeed, this  model can  accurately reproduce  the continuum shape observed in PG1247: the best fit gives \xddof$=325.5/365$.

To fit the \xmm\ and \nus\ data, a mildly ionized ($\log(\xi)\sim2.4$ erg cm s$^{-1}$), almost
Compton thick (\nh$\sim1.3\pm0.3\times10^{24}$ cm$^{-2}$) absorber is required. 
The covering fraction ($CF$) of the absorber is $CF\sim0.6$. Fig.~8 (left) shows the confidence contours for the column
density and the covering fraction. 
Adding a standard reflection component with $R=1$ has the only effect of reducing the covering factor to $CF\sim0.4$.

We note that the validity of {\it Xstar} is restricted to \nh$\leq1\times10^{24}$ cm$^{-2}$ in order to
avoid significant effects from Compton scattering (Kallman \& Bautista 2001),
even if the nominal \nh\ range of {\it ZXIPCF} reaches \nh$=5\times10^{24}$ cm$^{-2}$.
However, even imposing an \nh\ upper limit equal to \nh$=1\times10^{24}$ cm$^{-2}$ in our fit, does not  significantly change
the results for the other parameters or  the quality of the final best fit.

This model implies that the intrinsic luminosity of the source is a factor of $\sim2$ higher than the value observed by \xmm,
going from \lum$=8\times10^{45}$ \ergs\ to \lum$=1.5\times10^{46}$ \ergs. 
This, however, has a limited effect on the bolometric luminosity of the source, which is estimated to be \lbol$=1.5\times10^{48}$ \ergs\ 
from the optical continuum, given that at these luminosities the X-ray contribution is expected to be a small fraction of the total luminosity: 
the bolometric correction $k_{\rm bol}$ (defined as \lbol/\lum) is $\sim100$ already at \lbol$=10^{47}$ \ergs\ (Steffen et al. 2006, Lusso et al. 2012).

Interestingly, this model naturally produces absorption features at energies above the \feka\ line 
and could also produce significant \feka\ emission if the covering factor of the wind/outflow is large (i.e., P Cygni profile, see Nardini et al., 2015).
A similar feature is indeed observed in the \xmm\ spectrum of PG1247 (\nus\ does not cover the energy of interest), 
albeit at low significance ($\sim2\sigma$) and at energies slightly higher ($E_{\rm abs}\sim7.8$ keV rest frame) than 
the value expected for FeXXV-FeXXVI K$\alpha$ transitions ($6.70-6.97$ keV).
Therefore, if produced by FeXXV-FeXXVI, the observed feature would imply an outflow velocity 
of $v_{\rm out}\sim0.15c$, comparable to typical outflow velocities observed in local Seyferts (Tombesi et al. 2010).
However, the quality of the spectrum and the low significance of the feature do not allow us to investigate in detail the properties of this potential outflow.
Fig.~8 (right) shows the ratio (in rest-frame energy) between the data and a simple power law in the region of the \feka\ line.
We note that the absorption feature is seen at energies $E_{\rm abs}\sim2.4-2.6$ keV observed frame, and therefore just above 
the prominent edge in the \xmm\ mirror effective area at $E_{\rm edge}=2.2-2.3$ keV due to the Au M edge. 
This can produce some additional uncertainties in the instrumental calibration at these energies.  

If this model is correct, we have detected for the first time the high energy cutoff of the primary continuum emission in a high redshift, non-lensed quasar
(but see also Dadina et al. 2016, submitted, for the detection of a high-energy cutoff in a lensed QSO at z=$3.6$).
The low value measured may have important implications for the electron temperature of the corona (see  discussion is Sec.~6.2).

%======================================================
 \begin{figure*}
 \begin{center}
 \includegraphics[width=7cm]{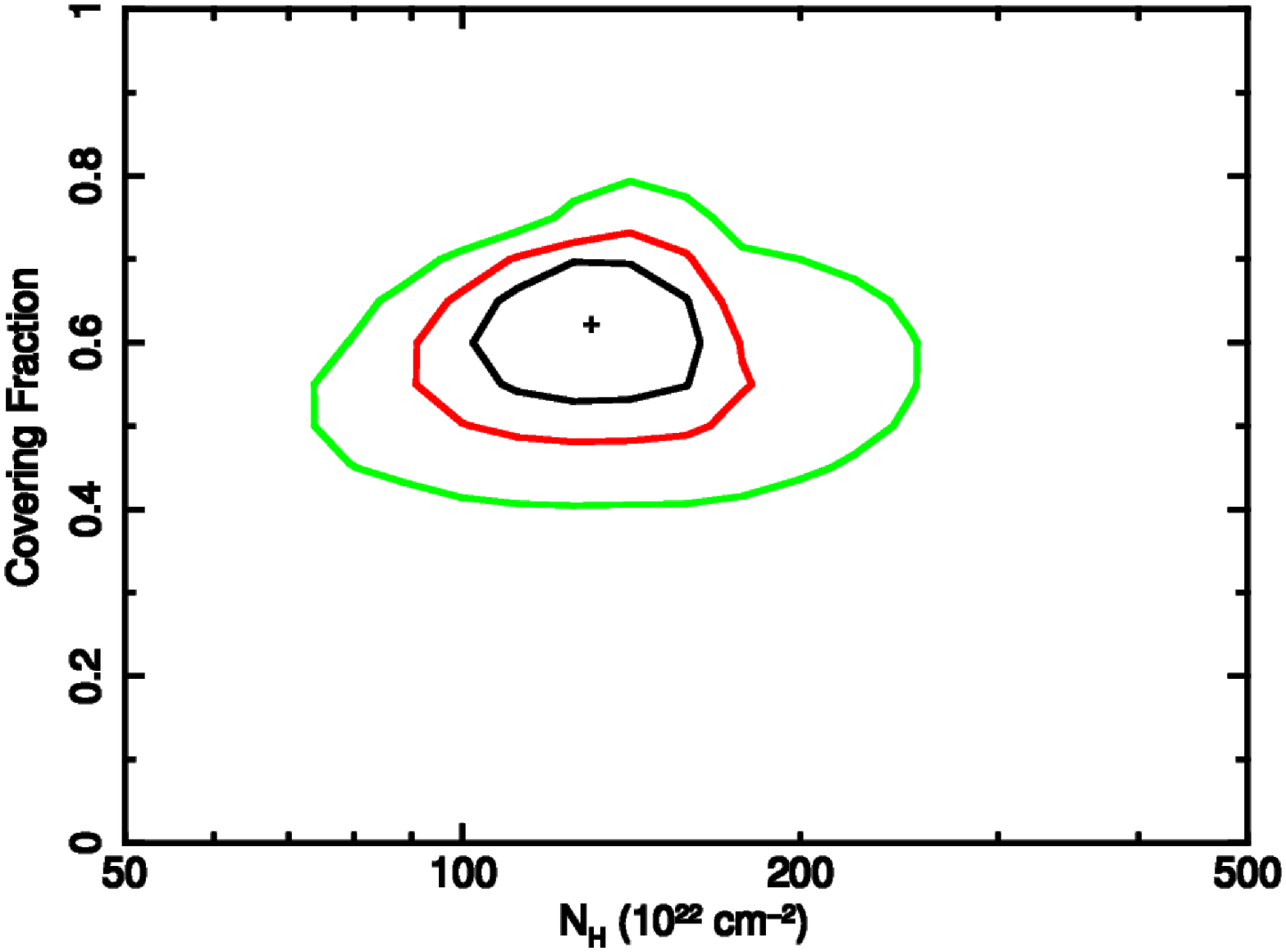}\hspace{1cm}\includegraphics[width=7cm]{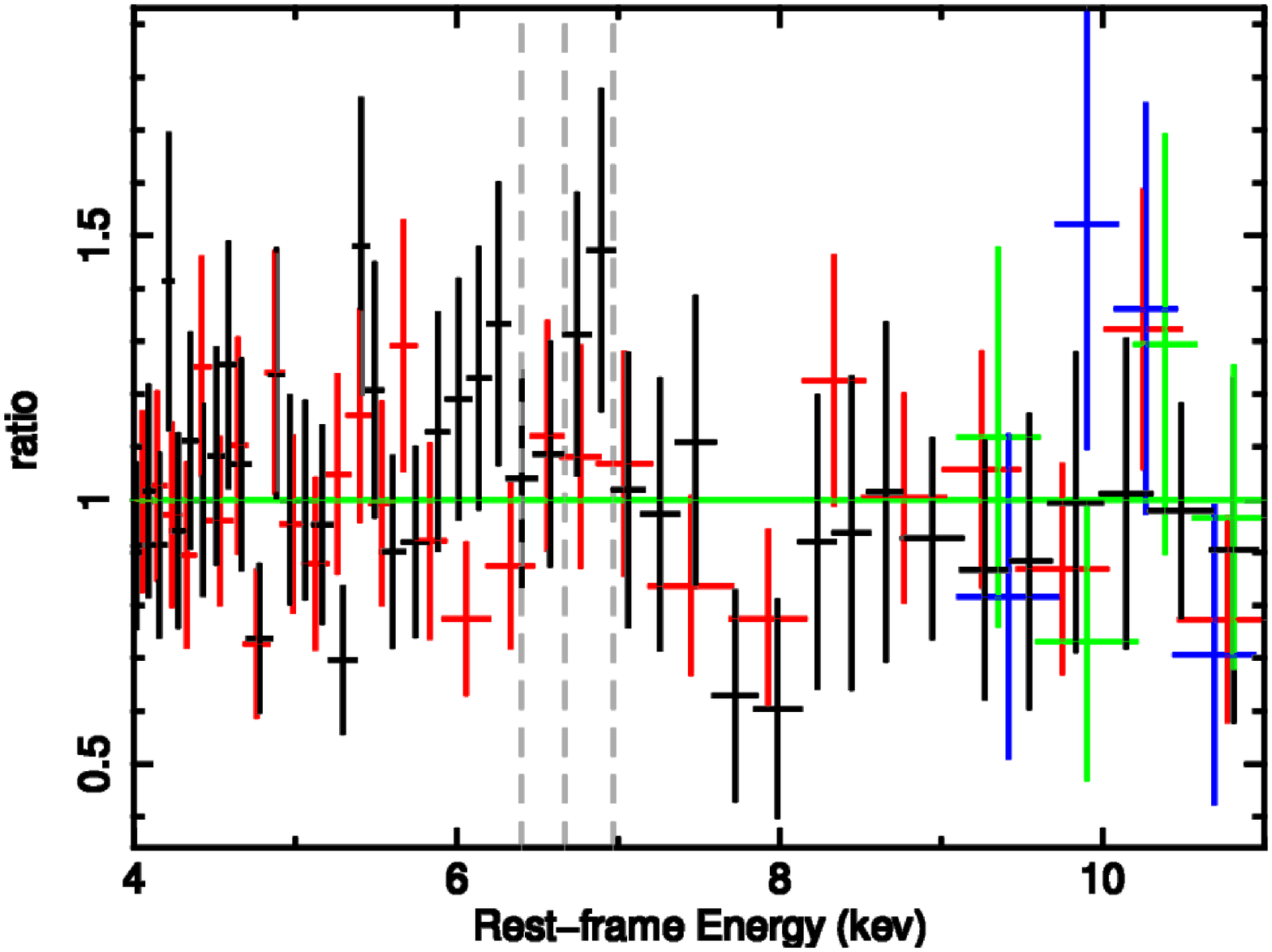}
 \caption{{\it Left:} Contour plot of the column density vs. covering fraction for model 3.
 {\it Right:} Rest-frame residuals in the \feka\ line region, after fitting with a simple power law. The gray dashed lines mark the expected neutral 
 Fe, FeXXV, and FeXXVI K$\alpha$ emission line energies. Symbols as in Fig.~6.
}
 \label{fig:selection}
 \end{center}
 \end{figure*}
%=======================================================

\subsection{Relativistic reflection} 

Finally, to test the alternative possibility that the reflection continuum is produced in the inner regions of the accretion disk, i.e., by ionized material
and under relativistic effects, we used the {\it RELXILL} model (Garcia et al. 2014), 
which is the convolution of the XILLVER reflection model (Garcia et al. 2010) for ionized accretion disk reflection, 
with the relativistic treatment of {\it RELCONV} (Dauser et al. 2010). 
In particular the {\it RELXILL$\_$LP} configuration assumes a simple lamp post geometry, i.e., a point-like hard X-ray source (the corona)
above the black hole that is irradiating the accretion disk (e.g., Matt et al. 1991, Dauser et al. 2013) and computes both the expected emissivity profile and the 
strength of the reflection self-consistently (model 4, Fig.~5 bottom right). 

The free parameters, in addition to the primary power-law photon index, high-energy cutoff, and normalizations,
are the height of the primary source ($h$), the BH spin ($a$), the inclination angle of the disk ($i$), and
the ionization parameter ($\xi$), defined as the ionization parameter of the accretion disk.
The inner radius of the accretion disk is fixed to the innermost stable circular orbit, which depends on the BH spin,
while the outer radius is frozen to 400 $r_{\rm G}$.

Interestingly the high-energy cutoff seems to be unconstrained for this model. This is not entirely due to the 
limited spectral quality of PG1247 and the larger number of free parameters for this model, but also to the fact that 
-- for a given primary power-law cutoff -- the reflection component computed by {\it RELXILL}  always has a sharper decline at high energies than the value computed by {\it PEXRAV} (see, e.g., Dauser et al. 2016). 
This means that the relativistic reflection model is able to reproduce the shape of the high-energy spectrum without the need of a cutoff.
Therefore, we fixed it to \ecut$=100$~keV, in order to be consistent with the measurements of the other parameters performed from the models discussed above.
The goodness of this fit is comparable with the fits obtained for the previous three models (\xddof$=333.8/365$).
The photon index is very soft ($\Gamma=2.26\pm0.04$) also for this model.

%======================================================
 \begin{figure*}
 \begin{center}
 \includegraphics[width=7cm]{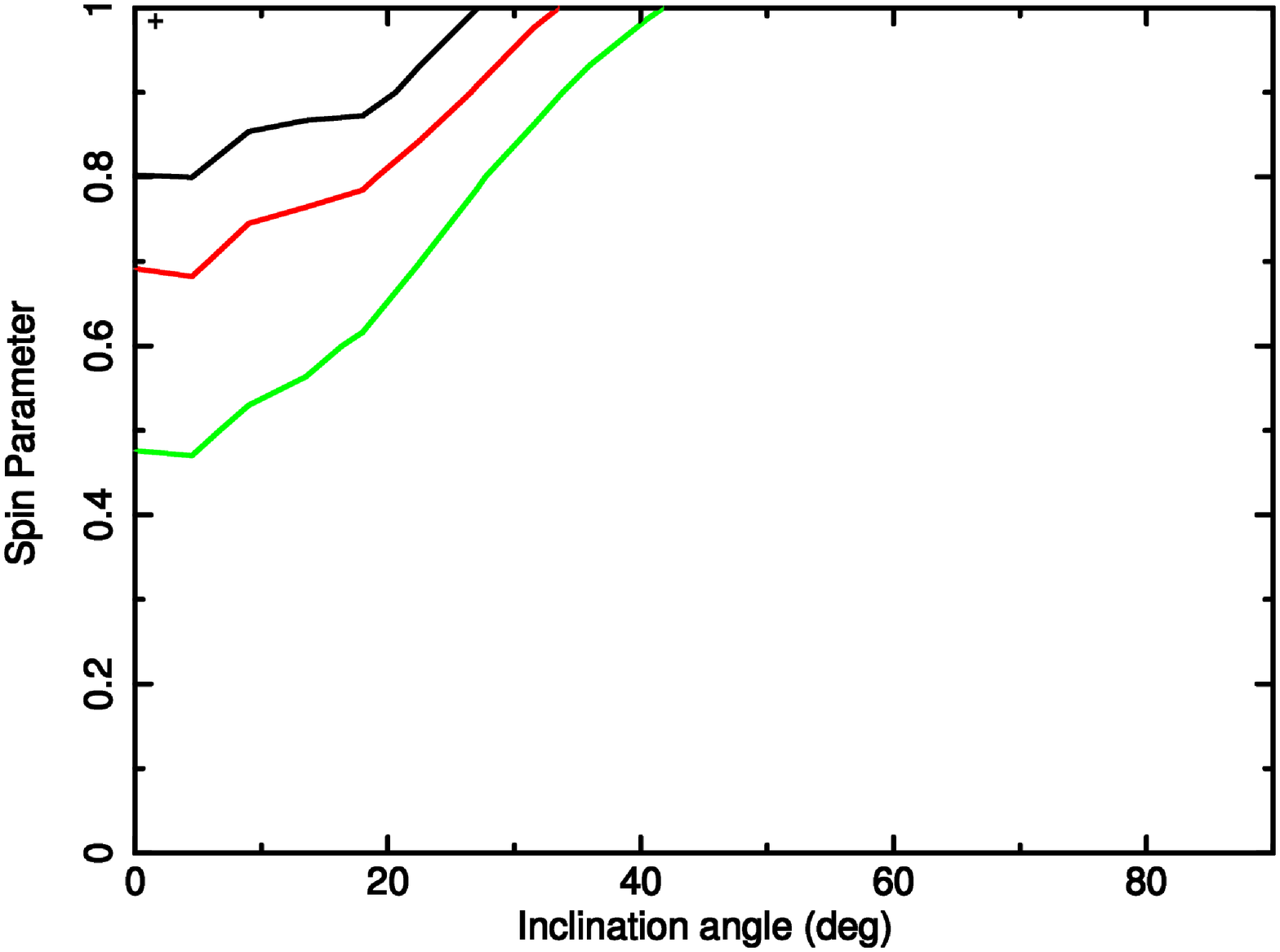}\hspace{1cm}\includegraphics[width=7cm]{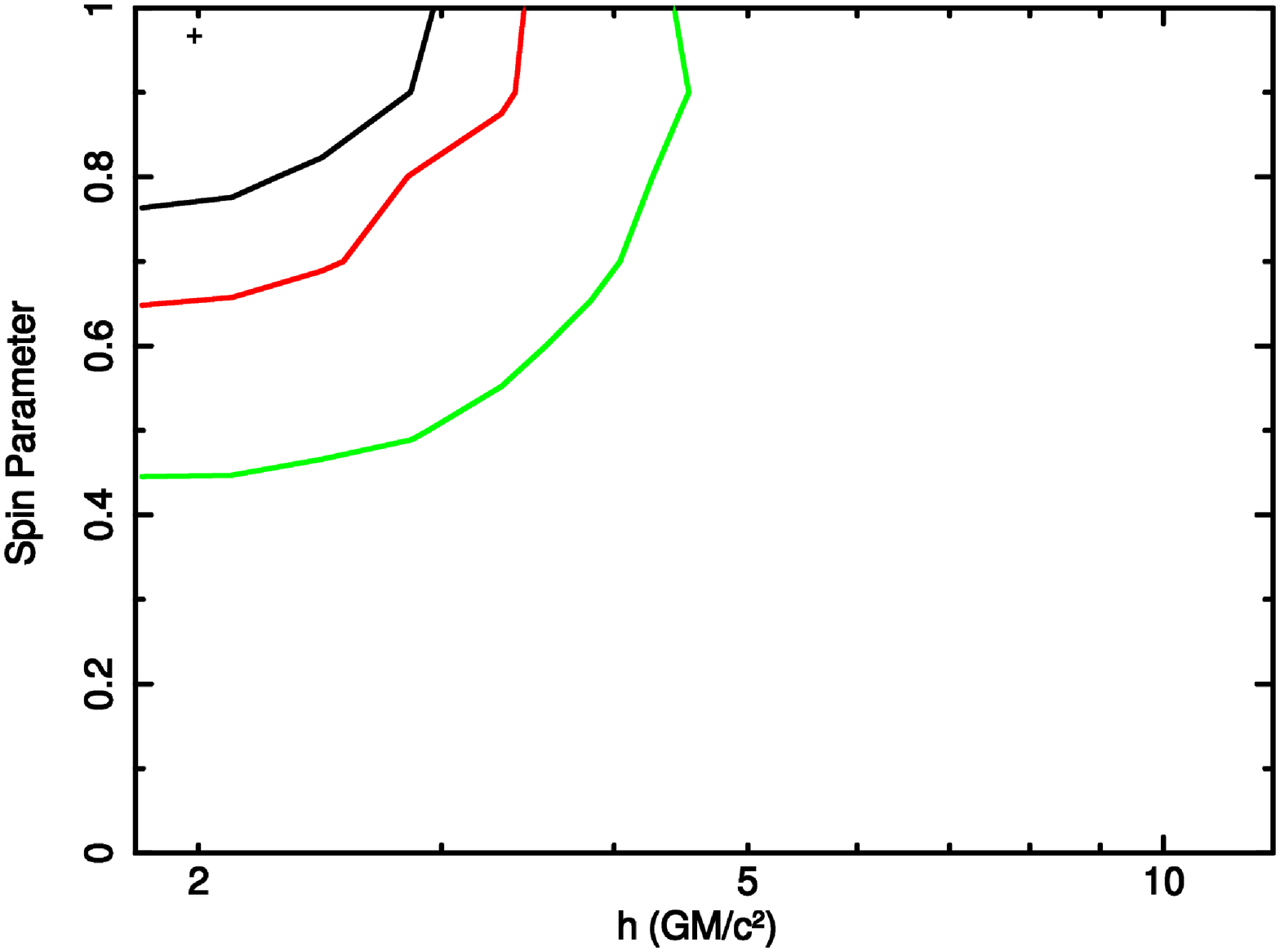}
 \caption{{\it Left:} Contour plot of the disk inclination angle vs. BH spin parameter for model 4.
 {\it Right:} Contour plot of the corona height $h$ vs. spin parameter for model 4.
 }
 \label{fig:relxill}
 \end{center}
 \end{figure*}
%=======================================================

The interesting aspect of this model is that the reflection fraction in this configuration is not a free parameter, but instead the reflection contribution is
computed self-consistently as the result of the geometry, size, and ionization state of the emitting corona.
The height of the corona, the BH spin, and the reflection fraction are indeed correlated in this model, 
in the sense that in order to produce a strong reflection fraction, both a low value of $h$ (i.e., the corona is close to the BH and hence to the disk) 
and a high value of the BH spin (i.e., the disk has a small inner radius) are  required (see Dauser et al. 2014).
Fig.~\ref{fig:relxill} (left) shows the inclination angle vs. BH spin parameter confidence contours:
to produce the strong reflection we see in the spectrum of PG1247, a nearly maximally rotating BH is required ($a>0.68$), 
observed at a small inclination angle ($\cos(i)>0.83$). 
Fig.~\ref{fig:relxill} (right) shows the confidence contours between the coronal height $h$ (in units of $r_{\rm G}$)
and the spin parameter. The height of the emitting region is constrained to be $h\leq3.5$ $r_{\rm G}$.
Finally, the ionization parameter is constrained to be $\log(\xi)=1.8\pm0.5$ erg cm s$^{-1}$; 
i.e., the disk must be moderately ionized in order to produce the strong Compton hump, 
while a more ionized reflecting medium would produce a steeper reflection continuum. 
We note that a small height of the corona ($h< 10$ $r_{\rm G}$) for a maximally rotating BH ($a>0.9$), observed at small inclination angles ($i<30^{\circ}$),
has also  been estimated  in a few local Seyferts through spectral-timing analysis and reverberation techniques (e.g., Cackett et al. 2014; see Uttley et al. 2014 for a review), 
while evidence is mounting that maximally rotating SMBHs may be common among AGN at low and high redshift (Walton et al. 2013; 
Reis et al. 2014; Reynolds et al. 2014), with the caveat that high spin means high accretion efficiency (Vasudevan et al. 2015) 
so the brightest objects (for which measuring the BH spin is feasible) in a population of objects with similar accretion properties will be those with high spin.

\section{Discussion}

Several aspects of the X-ray broad-band spectrum of PG1247 are in agreement with the possibility that this high-redshift QSO is 
accreting close to or above the Eddington limit, independent of the model adopted to reproduce the overall spectral shape:

\begin{itemize}

\item The photon index is very soft: $\Gamma\sim2.3-2.4$ for all models. 
These values are consistent with those expected for a nearly Eddington accreting SMBH, given the relation observed in individual, variable, local AGN 
(e.g., Perola et al. 1986, Vaughan \& Edelson 2001, Vignali et al. 2008, Puccetti et al. 2014), and in samples of low- and high-redshift AGN (Shemmer et al. 2008, 
Risaliti et al. 2009, Brightman et al. 2013). The caveat is that the parameter space above the Eddington limit is currently unexplored.
Unfortunately, owing to the limited data quality and the lack of simultaneity between the \xmm\ and \nus\ data,
we cannot test whether PG1247 shows the {\it softer when brighter} behavior within the flux variation observed 
between the \xmm\ and \nus\ data.\\
\item The Fe emission line is most likely produced by ionized gas in the accretion disk (the rest-frame energy of 6.4 keV is ruled out at 90\% c.l.),
as observed in the stacked spectrum of large samples of AGN accreting close to the Eddington limit (Iwasawa et al. 2012).
The EW of the line (EW$=168_{-150}^{+198}$~eV) is consistent, within the errors, with the value expected taking into account 
the high reflection factor derived from {\it PEXRAV} (possibly due to variability, see below) and the X-ray Baldwin effect 
(EW$\sim20-30$ eV intrinsic, to be rescaled by a reflection parameter $R=3.8$).\\
\item The residuals above the expected energies of \feka\ suggest the presence
of an outflowing disk wind, as predicted by super-Eddington accretion models (e.g., Zubovas \& King 2012),
and observed in a handful of high-redshift QSOs accreting close to Eddington (Chartas et al. 2003, Lanzuisi et al. 2012, Vignali et al. 2015). 
The presence of ionized material surrounding the SMBH is indeed one of the characteristics of highly accreting BHs
(see, e.g., Ballantyne et al. 2011).
\end{itemize}

\subsection{Reflection and accretion}

The intensity of the reflection component is usually thought to be anti-correlated with the flux state, and therefore \edd,
so that sources accreting closer to the Eddington limit show lower levels of reflection (see, e.g., Fabian et al. 2012, Keek \& Ballantyne 2015).
Therefore, the super-Eddington nature of PG1247 and the exceptional reflection component observed in its spectrum seem in tension.
However, with the exception of model 1, which requires an unphysical value of the reflection fraction ($R>1$), 
each of the other models discussed in Sec. 5 offers an explanation, even though the conclusions are strongly model-dependent.

In PG1247, the SMBH is on average accreting close to or above Eddington, while the \xmm\ observation of 2003 can be considered a drop in the X-ray flux of a factor of $\sim2$  with respect to the average, and a factor of $\sim3$ with respect to the \rosat\ measurements.
The cold reflection scenario is therefore able to explain the exceptionally high reflection component in terms of variability (model 2): 
the source was a factor of $\sim3$ fainter at the time of the \xmm\ observation than during the \rosat\ observation. 
In this scenario the super-Eddington accretion rate has nothing to do with the 
presence of a strong reflection component, which is only a light echo of the continuum level as it was 1200 days before the \xmm\ observation.
A reflection fraction $R=3-4$ in the \xmm\ data is therefore in agreement with this scenario, while the reflection fraction for the \nus\ data
in model 2 is consistent with $R=1$ within the errors. 

We note that the value $R=1$ in the {\it PEXRAV} model is defined relative to the reflection produced by an infinite slab of cold material illuminated by the corona.
If instead the reflector is the inner wall of the torus, we do not expect a reflection as high as $R=1$ (the total solid angle seen by the reflector must be $<2\pi$ sr).
However, if we imagine a torus with a small half opening angle and large height (e.g., the one described in Ikeda et al. 2009), 
$R$ could be close to $\sim1$. 
Furthermore, if the reflection component is produced by cold material at the inner edge of the 
obscuring torus, the time delay between continuum and cold reflection variability would imply a radius of the torus inner edge of $\sim1$ pc, 
roughly in agreement with the expected sublimation radius $r_{in}=1-4$ pc (Barvainis 1987; Kishimoto et al. 2007)
given the UV luminosity of $\lambda L_{\lambda}(1350\AA)\sim4\times10^{47}$ \ergs\ (Trevese et al. 2014) and with the torus half-light radius 
expected from the bolometric luminosity, i.e., $1-10$ pc at $10^{48}$ \ergs\ (Burtscher et al. 2013).

The ionized absorber scenario (model 3) is able to explain the broad-band shape observed in PG1247 
without the need of any reflection component. The spectral curvature is due to the effect of a mildly ionized ($\log\xi\sim2.4$ erg cm s$^{-1}$) dense absorber (\nh$\sim1\times10^{24}$ cm$^{-2}$) 
with a covering factor of $CF\sim0.6$ superimposed on a simple power law with a high-energy cutoff \ecut$\sim100$ keV. 
The addition of a standard cold reflection with $R=1$ would reduce the required covering factor to $CF\sim0.4$.
In this case, the link between the shape of the X-ray spectrum of PG1247 and its extreme accretion properties would be in the presence of 
ionized, outflowing material close to the SMBH, a clear prediction of super-Eddington accretion theories 
that needs to be investigated with higher $S/N$ spectra in the region of the \feka\ emission line.  

Finally, the relativistic reflection scenario (model 4) is able to explain the strong reflection component with an extreme geometry of the system,
i.e.,  a rather small inclination angle ($i<30^{\circ}$), a small height of the corona ($h<3.5$ $r_{\rm G}$), and a maximally rotating BH ($a\sim0.97$). 
The face-on geometry required  to explain the strong reflection may actually affect the SMBH mass measurements,
in particular requiring an increase in both the SE and RM mass estimates by a factor of up to 5 for very small angles (Pancoast et al. 2014).
This would reduce the Eddington ratio of PG1247 by the same amount.

\subsection{\ecut\ and the properties of the corona}
%======================================================
 \begin{figure*}
 \begin{center}
 \includegraphics[width=6cm]{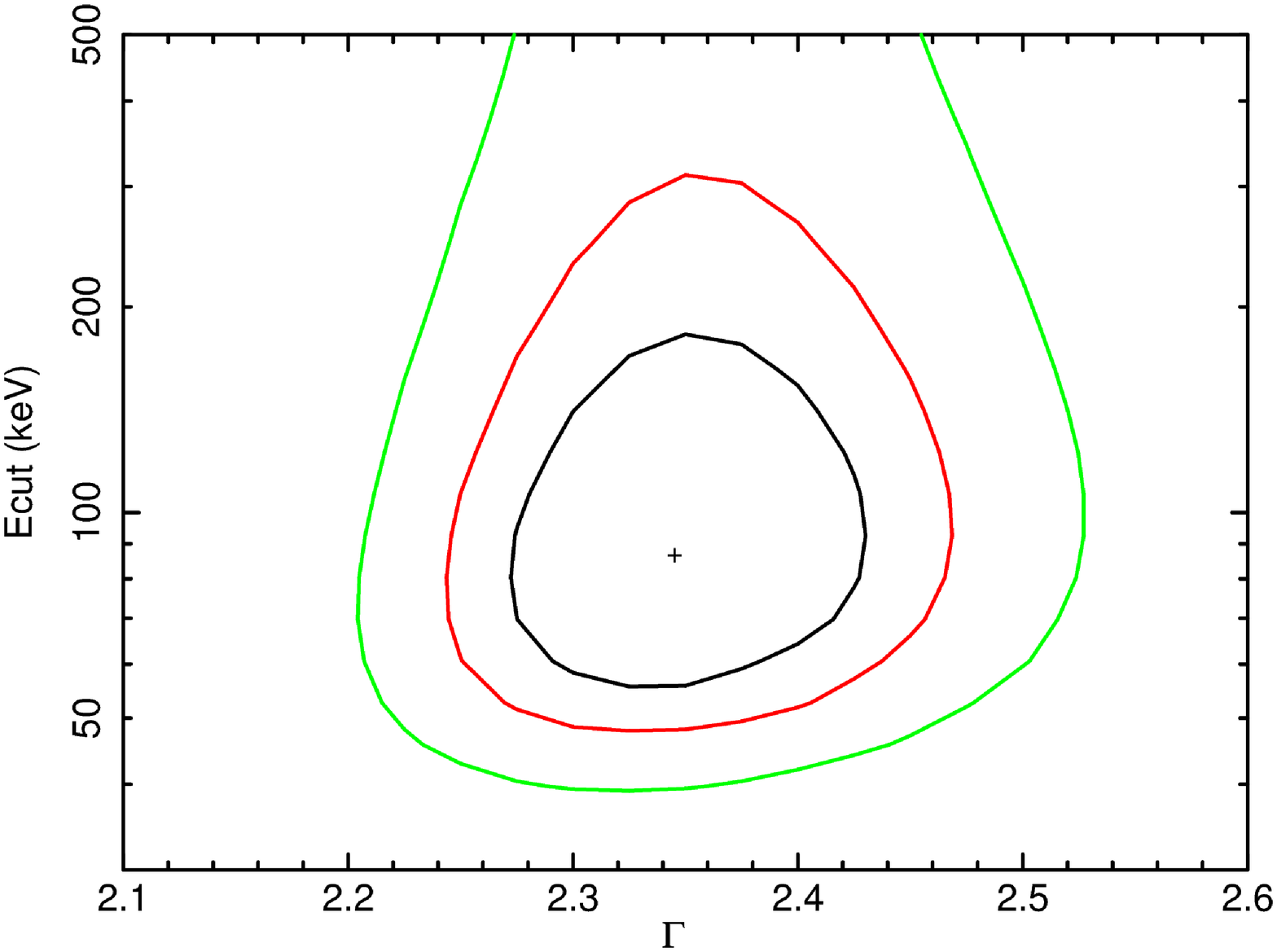}\hspace{0.2cm}\includegraphics[width=6cm]{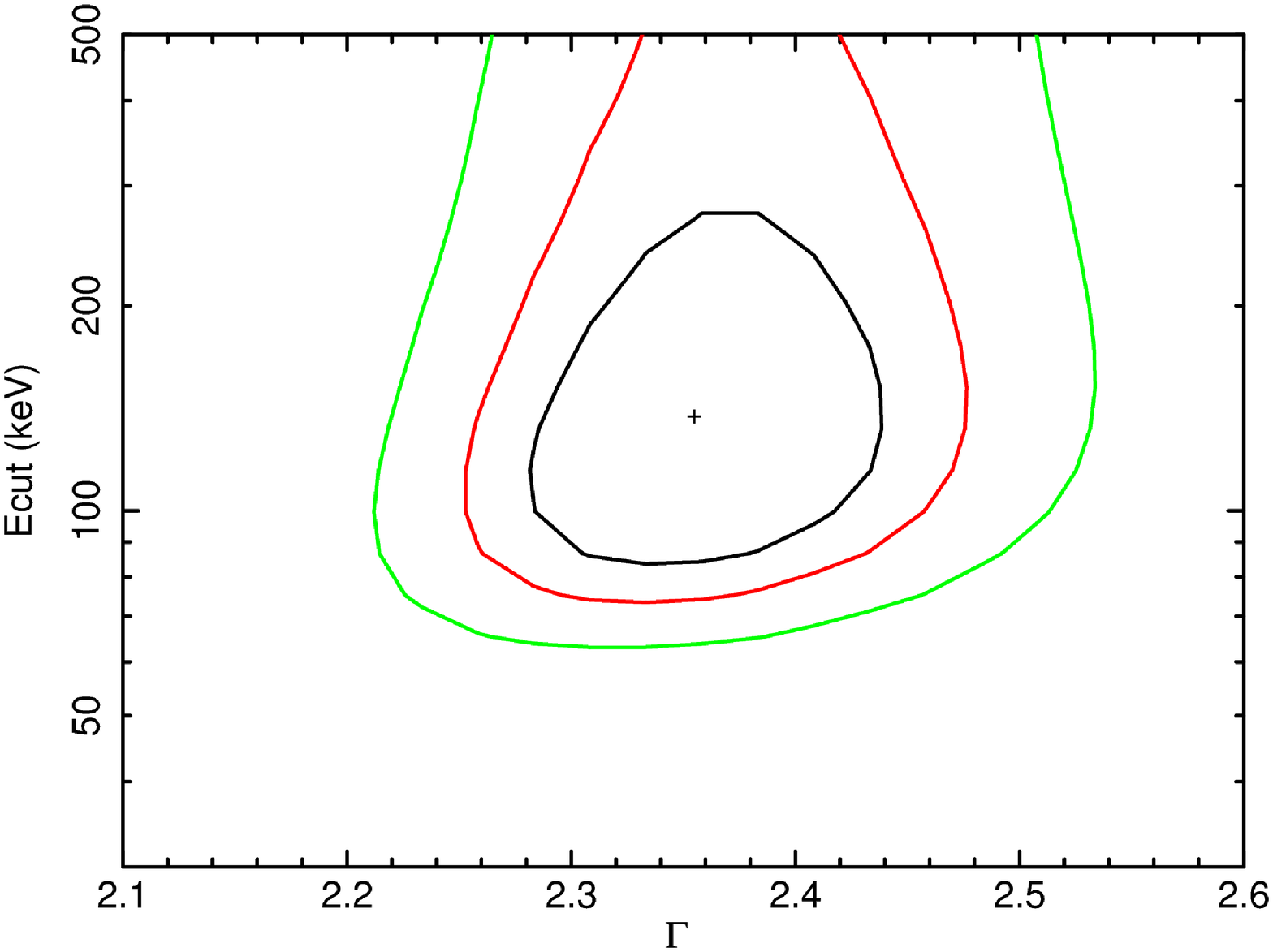}\hspace{0.2cm}\includegraphics[width=6cm]{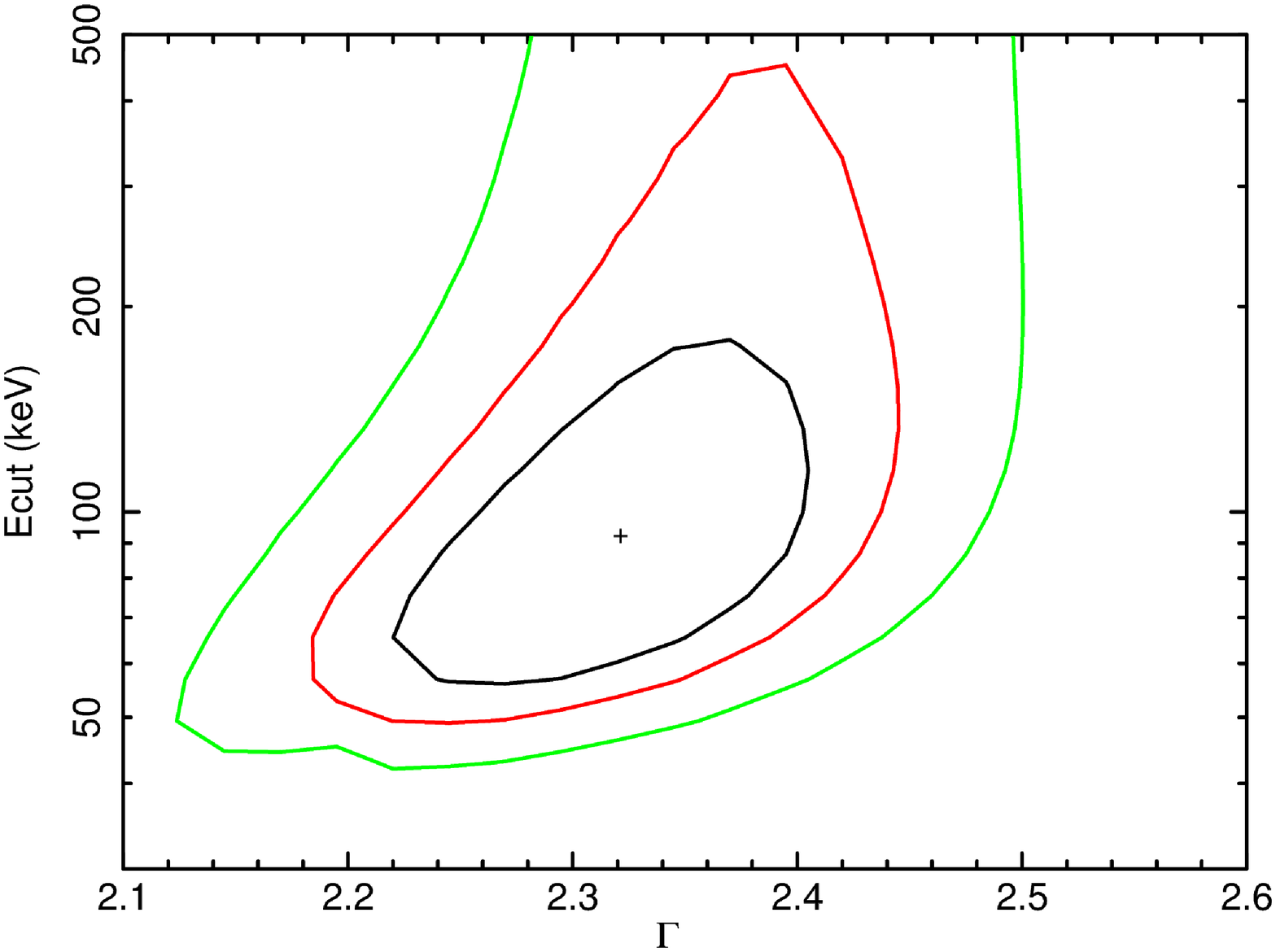}
 \caption{Confidence contours (at 68, 90, and 99\% c.l.) of  $\Gamma$ vs. \ecut\ (rest frame) parameters for model 1 (left),  model 2 (center), and model 3 (right).}
 \label{fig:cont}
 \end{center}
 \end{figure*}
%=======================================================

The confidence contours for $\Gamma$ and \ecut\ measured from the different models are shown in Fig. 10.
The results on \ecut\ obtained from model 1 (left panel) are not considered here since, as discussed above, this model must be considered unrealistic.
The best-fit \ecut\ measured with model 2 (central panel) is \ecut $\sim$ 160~keV, but it is only loosely constrained to be \ecut\simlt600~keV.
For model 3 instead, we clearly constrain, for the first time at such high redshift, a rather low cutoff (right panel).
Finally, for model 4 the \ecut\ is unconstrained and therefore fixed in the spectral fit.

The high-energy cutoff measured in model 3 is consistent with the average \ecut\ observed at low redshift (e.g., Malizia et al. 2014: $\langle$\ecut$\rangle=128$ keV, 
with standard deviation $\sigma=46$~keV.
This rather low value would imply a low plasma temperature of the Comptonizing region ($kT_e=$\ecut$/2$ for $\tau\simlt1$ and large \edd).
Indeed, using Eq. 1 of Petrucci et al. (2001) and the derived \ecut\ and $\Gamma$ values, we can derive the optical depth of the corona.
Using the results from model 3, the optical depth is $\tau\sim1.3$, i.e., the corona is optically thick.
We stress, however, that -- as shown in Petrucci et al. (2001) -- the $kT_e$ values derived from {\it PEXRAV} 
are in general underestimated, while the $\tau$ is overestimated with respect to a realistic,
anisotropic Comptonization model, and therefore these results must be taken with caution.

We stress that the result obtained for model 3 is, to our best knowledge, the first detection of a high-energy cutoff at such high redshift.
The rather low temperature derived from the high-energy cutoff and the high compactness of the corona 
($\ell\simgt10$, with $\ell$ defined as $\ell=L \sigma_{T}/R m_e c^3$, Guilbert, Fabian \& Rees 1983)
are in agreement with our current understanding of the heating and thermalization mechanisms operating in the corona where 
large powers are dissipated in the physically compact regions surrounding the SMBH (Fabian et al. 2015).

PG1247 is, therefore, a clear example of how \nus\ is opening up the possibility of measuring the fundamental parameters of coronal emission 
($\ell$, $\tau$, and $kT_e$) beyond the local Universe, thanks to sensitive hard X-ray observations.
However, a campaign of deep \xmm\ observations, coupled with simultaneous hard X-ray data from \nus, 
are required in order to understand which of the models discussed in Sec.~5 best represents the observed X-ray spectrum of PG1247,
e.g., distinguishing between cold and relativistic reflection or clearly detecting absorption troughs associated with ionized obscuring gas.

\begin{acknowledgements}

We thank the anonymous referee for constructive comments that have helped us to improve the quality of the paper.
GL thanks F. Gastaldello for useful insights on \nus\ background issues,
E. Dalessandro for advice about HST data, and O. Shemmer for help with the \swift\ data. 
GL acknowledges financial support from the CIG grant ``eEASY'' n. 321913 and from ASI-INAF 2014-045-R.0 and ASI/INAF I/037/12/0–011/13 grants.
ACF acknowledges support from ERC grant 340442.
FEB acknowledges support from CONICYT-Chile (Basal-CATA PFB-06/2007, FONDECYT Regular 1141218, ``EMBIGGEN'' Anillo ACT1101), 
the Ministry of Economy, Development, and Tourism's Millennium Science Initiative through grant IC120009, 
awarded to The Millennium Institute of Astrophysics, MAS.
WNB and BL acknowledges support from Caltech NuSTAR subcontract 44A-1092750.
This work made use of data from the \nus\ mission, a project led by the
California Institute of Technology, managed by the Jet Propulsion
Laboratory, and funded by NASA.
This research also made use of the NuSTAR Data Analysis
Software (NuSTARDAS) jointly developed by the ASI Science
Data Center (ASDC, Italy) and the California Institute of
Technology (USA)

\end{acknowledgements}

%-------------------------------------------------------------------

\end{document}